\documentclass[usenatbib]{mnras}
\usepackage{IEEEtrantools} %
\usepackage{amsmath} %
\usepackage{amssymb} %
\usepackage{bm} %
\usepackage{upgreek} %
\usepackage[pdftex]{graphicx} %
\usepackage{multirow}
\usepackage{color}

\def\apjl{ApJ}
\def\apjs{ApJS}
\def\ao{Appl.~Opt.}

\def\aap{A\&A}

\newcommand{\expo}[1]{\mathrm{e}^{#1}} 

\newcommand{\feh}{\text{[Fe/H]}}

\newcommand{\lna}[1]{\text{OI\,#1\,nm}}
\newcommand{\lnf}[1]{\text{[OI]\,#1\,nm}}

\newcommand{\teff}{T_{\text{eff}}}
\newcommand{\lgg}{\log g}
\newcommand{\lggu}{\log\left(g / \mathrm{cm\,s^{-2}}\right)}
\newcommand{\ud}{\mathrm{d}}
\newcommand{\lgeps}{\log\epsilon_{\text{O}}}
\newcommand{\corr}[2]{\Delta^{\text{#1}}_{\text{#2}}}
\newcommand{\lgt}{\log\tau_{500}}
\newcommand{\lgr}{\log\tau_{\mathrm{R}}}

\newcommand{\multi}{\textsc{Multi}}
\newcommand{\mtd}{\textsc{Multi3D}}

\newcommand{\stagger}{\textsc{Stagger}}
\newcommand{\atmo}{\textsc{Atmo}}
\newcommand{\scate}{\textsc{Scate}}
\newcommand{\marcs}{\textsc{Marcs}}

\newcommand{\atlasnine}{\textsc{Atlas9}}
\newcommand{\tlusty}{\textsc{TLusty}}
\newcommand{\synspec}{\textsc{SynSpec}}
\newcommand{\detail}{\textsc{Detail}}

\newcommand{\eqn}[1]{\text{Eq.~\ref{#1}}}
\newcommand{\sect}[1]{\text{Sect.~\ref{#1}}}
\newcommand{\fig}[1]{\text{Fig.~\ref{#1}}}
\newcommand{\tab}[1]{\text{Table~\ref{#1}}}

\newcommand{\term}[4]{#1#2 $^{#3}$#4}
\newcommand{\tripletlo}{\term{3}{s}{5}{S}}
\newcommand{\tripletup}{\term{3}{p}{5}{P}}
\newcommand{\oground}{\term{2}{p}{3}{S}}

\newcommand{\markcorr}[1]{#1}

\newcommand{\kms}{\mathrm{km\,s^{-1}}}
\newcommand{\expect}[1]{{\rm I\kern-.3em E}\left[#1\right]}
\newcommand{\sh}{S_{\mathrm{H}}}

\title[3D non-LTE OI line formation]{Non-LTE oxygen line formation in 3D hydrodynamic model stellar atmospheres}
\author[A.~M.~Amarsi, M.~Asplund, R.~Collet, and J.~Leenaarts]{A.~M.~Amarsi$^{1}$\thanks{E-mail: anish.amarsi@anu.edu.au}
, M.~Asplund$^{1}$, R.~Collet$^{1}$, and J.~Leenaarts$^{2}$\\
$^{1}$ Research School of Astronomy and Astrophysics,
Australian National University, ACT 2611, Australia\\
$^{2}$ Institute for Solar Physics, 
Stockholm University, SE-106 91 Stockholm, Sweden}
\begin{document}

\date{Accepted 2015 November 03.  Received 2015 November 03; in original form 2015 September 07}
\pagerange{\pageref{firstpage}--\pageref{lastpage}} \pubyear{---}

\maketitle 
\label{firstpage}
\begin{abstract}
The \lna{777}~lines are among the most commonly used
diagnostics for the oxygen abundances
in the atmospheres of FGK-type stars.
However, they form in conditions that are far from 
local thermodynamic equilibrium (LTE).
We explore the departures from LTE of atomic oxygen,
and their impact on OI lines, across the 
\stagger-grid of three-dimensional hydrodynamic model
atmospheres.
For the \lna{777}~triplet 
we find significant departures from LTE.
These departures are larger in stars with 
larger effective temperatures,
smaller surface gravities,
and larger oxygen abundances.
We present grids of predicted 3D non-LTE based 
equivalent widths 
for the \lna{616}, \lnf{630}, \lnf{636}, and \lna{777}~lines,
as well as abundance corrections to 1D LTE based results.
\end{abstract}
\section{Introduction}
\label{introduction}

As the most abundant of the metals,
oxygen plays an important role
in many areas of astrophysics.
It has, for example,
a large impact on stellar structure and evolution
due to its large opacity contribution 
and the importance of the CNO nucleosynthesis cycle
\citep[e.g.][]{2012ApJ...755...15V},
to the extent that the stipulated errors
in the ages of metal poor 
stars and globular clusters inferred from
theoretical evolutionary tracks
can be dominated by uncertainties
in the oxygen abundances
\citep[e.g.][]{2003Sci...299...65K,2013ApJ...765L..12B}.
Oxygen can also have a large impact on
the chemistry of exoplanetary atmospheres
\citep{2012ApJ...758...36M},
which is driving interest in determining 
the carbon-to-oxygen ratio in exoplanet host stars
\citep[e.g.][]{2011ApJ...735...41P,2014A&amp;A...568A..25N}.
Finally, since oxygen abundances can be measured
using different diagnostics in a variety of sites
and down to very low metallicities
\citep[e.g.][]{2012EAS....54.....S_short},
they are key tracers of the chemical
evolution of our galaxy
\citep[e.g.][]{1979ApJ...229.1046T,1997ARA&amp;A..35..503M}.

Abundances of oxygen, and indeed of all chemical species,
in the atmospheres of stars 
cannot be measured directly.
They must instead be inferred by matching
observations to models.
Inadequacies in the models
of stellar atmospheres and spectrum formation
can cause the extracted abundances
to be unrealistic. 

In particular, large systematic errors
can arise by 
imposing local thermodynamic equilibrium (LTE).
It has long been known
\citep{1957ApJ...125..260T,1959AnAp...22..499P,
1966IAUS...26..207J,1967SoPh....1...27D}
that the photospheres of late-type stellar atmospheres
show significant departures from LTE
since the densities of electrons and neutral hydrogen
can become so small that 
energy partitioning according to Boltzmann statistics
is no longer maintained 
\citep[e.g.][]{1973ARA&amp;A..11..187M}.
This has direct implications
on the strengths of the lines 
which form there, and thus on the inferred abundances.

Errors can also arise by
employing one dimensional (1D) hydrostatic model atmospheres,
which neglect atmospheric inhomogeneities,
and typically predict, for a given 
effective temperature and surface gravity,
temperature stratifications 
that are too shallow \markcorr{in the upper atmospheric layers}~at 
low \feh \citep[e.g.][]{1999A&amp;A...346L..17A}~and
too steep at \markcorr{solar metallicities
\citep[e.g.][]{2013A&amp;A...554A.118P}~compared}
to the mean stratification of their 3D counterparts.
These effects can couple to the 
departures from LTE
to affect the line formation in complicated ways,
and in general do not average out  
\citep[e.g.][]{2005ARA&amp;A..43..481A}. 
The inadequacies of 1D hydrostatic models 
stem from their intrinsic inability
to realistically describe the 
transport of energy by convection processes 
and from the lack of self-consistent
atmospheric velocity fields
\citep[e.g.][]{2007A&amp;A...469..687C}.
In that respect, ab initio 3D hydrodynamic models 
\citep[e.g.][]{2009LRSP....6....2N,
2012JCoPh.231..919F,2013A&amp;A...557A..26M}
are far preferable.

In FGK-type stars the \lna{777}~triplet lines
are among the most commonly used oxygen
abundance diagnostics,
as they are typically strong and free of blends.
However, these lines form in conditions
that depart from LTE.
Non-LTE effects were first noted in these lines
by \citet{1968SoPh....5..260A},
and later critically studied with multi-level
non-LTE radiative transfer techniques by  
\citet{1974A&amp;A....31...23S}.
For a given abundance the lines were
found to be significantly stronger in non-LTE,
a conclusion that has been supported by numerous
subsequent studies
\citep[e.g.][]{1979A&amp;A....71..178E,
1993A&amp;A...275..269K,
1999A&amp;A...350..955G,
2003A&amp;A...402..343T,
2009A&amp;A...500.1221F}.
Studies of non-LTE effects in three dimensional models
have hitherto been restricted to the Sun
\citep{1995A&amp;A...302..578K,2004A&amp;A...417..751A,
2009A&amp;A...508.1403P,2013MSAIS..24..111P,
2015arXiv150803487S};
such studies were critical for the recent downwards revision of 
over 0.2 dex of the canonical solar oxygen abundance 
\citep[cf.][]{2004A&amp;A...417..751A,
2009ARA&amp;A..47..481A,2008A&amp;A...488.1031C}.

Other permitted atomic oxygen lines
(in particular the \lna{615.8}~line)
see less use on account of
their weakness,
the presence of blends
and/or the lack of accurate atomic data.
They are less susceptible to non-LTE effects
than the \lna{777}~lines 
\citep[e.g.][]{2004A&amp;A...417..751A,2008A&amp;A...488.1031C}.
The forbidden 
\lnf{630.0}~and \lnf{636.3}~lines 
are a popular alternative \citep[e.g.][]{1978MNRAS.182..249L}
and have the tremendous advantage 
of forming essentially in LTE
\citep[e.g.][]{2004A&amp;A...417..751A}.
However, they are usually weak, 
sensitive to the model atmosphere
\citep[e.g.][]{2002A&amp;A...390..235N},
and severely blended \citep[e.g.][]{1978MNRAS.182..249L,
2001ApJ...556L..63A}.

Molecular oxygen lines, particularly OH 
electron excitation lines in the UV,
are also popular
\citep[e.g.][]{1998ApJ...507..805I,1999AJ....117..492B}.
However, to obtain accurate abundances 
from these lines, they must be modelled in 3D, 
owing to the steep temperature sensitivity of
molecule formation:
the abundances inferred from 1D models
can be over 0.5 dex too large
\citep{2001A&amp;A...372..601A,2007A&amp;A...469..687C}.
Multi-level non-LTE analyses
of OH lines have not yet been attempted,
although an exploratory study by \citet{2001A&amp;A...372..601A},
who used 1D models 
and employed the two-level approximation,
found that the non-LTE abundances were 
lowered by as much as 0.25 dex.
Non-LTE analyses of these lines
are limited by the lack of accurate collisional rate coefficients
and by the computational resources that they demand;
thus, full 3D non-LTE analyses of 
molecular oxygen lines still
await to be tackled. 

In this paper we detail
recent 3D multi-level non-LTE atomic oxygen line formation
calculations across a grid of 3D hydrodynamic 
\stagger~stellar model atmospheres.
After describing the simulation setup in \sect{method},
we present the results, including
a description of the non-LTE mechanism, the 
3D effects, and abundance errors, in \sect{results}.
We focus our discussion on the \lna{777}~lines,
although we discuss abundance errors
for the \lna{615.8}, \lnf{630.0}~and \lnf{636.4}~lines as well.
We compare our results with those
from related studies in \sect{comparison}
before summarising in \sect{conclusion}.

\section{Method}
\label{method}

\subsection{Overview}
\label{methodoverview}
We briefly review the governing equations of non-LTE radiative transfer
\citep[e.g.][]{2003rtsa.book.....R,2014tsa..book.....H}.
Non-LTE analyses typically assume 
statistical equilibrium,
\phantomsection\begin{IEEEeqnarray}{rCl}
\label{rateequation}
	\displaystyle\sum\limits_{j\neq i}\left(n_{j}\,P_{j\,i}-n_{i}\,P_{i\,j}\right) &=& 0 \, ,
\end{IEEEeqnarray}
where the transition rate from
energy level $i$ to energy level $j$
(which have corresponding populations
$n_{i}$ and $n_{j}$)
can be split into a
radiative component and a collisional component:
$P_{i\,j}=R_{i\,j}+C_{i\,j}$.
This equation states that in equilibrium, the 
flow into an arbitrary 
energy level is balanced by the flow out of it:
$\frac{\ud n_{i}}{\ud t}=0$.
The system of equations is closed by conserving
the total number density: $\displaystyle\sum\limits_{i}n_{i}=N$.
In stellar photospheres, particle velocities 
have to a good approximation
Maxwell-Boltzmann distributions
\citep[due to the efficiency of electron collisions in 
thermalising the medium; 
see][and references therein]{2014tsa..book.....H}.
\markcorr{Thus, collisions tend to drive
levels to be in relative LTE.
In the special case where collisional rates dominate over all of the
radiative rates, LTE is reproduced.} 

To obtain the radiative rates,
the radiation field must first be determined by 
solving the radiative transport equation
for the specific intensity $I_{\nu}$ along a ray, 
\phantomsection\begin{IEEEeqnarray}{rCl}
\label{transportequation}
	\frac{\ud I_{\nu}}{\ud s} &=& \alpha_{\nu}\,\left(S_{\nu}-I_{\nu}\right)\, ,
\end{IEEEeqnarray}
where $s$ is the path length
\citep[e.g.][]{2003rtsa.book.....R}. 
The linear extinction coefficient
$\alpha_{\nu}$ and
source function $S_{\nu}$
are functions of the populations,
making the problem strongly non-linear.
\markcorr{Radiative transfer couples different parts of the
atmosphere together, making the problem
a non-local one.}
Through the radiation field, all levels
are in principle coupled to each other,
everywhere in the atmosphere.

\subsection{Code description}
\label{methodcode}

The system of equations discussed 
in \sect{methodoverview} was solved
iteratively by the MPI-parallelised
domain-decomposed 3D non-LTE radiative transfer code 
\mtd~\citep{2009ASPC..415...87L}. Details of
the code can be found in that paper and
references therein; 
we briefly discuss some of its important aspects,
as well as recent changes made to the code.

\mtd~accepts 3D model atmospheres
defined on a Cartesian grid that is periodic
and has equidistant spacing in the two horizontal dimensions.
In the vertical direction the grid can be non-equidistant. 
The model atmospheres specify the 
gas temperature $T$, gas density $\rho$ 
and electron number density $n_{\text{e}}$.
The populations of all species not
treated in non-LTE are computed from these three thermodynamic
variables assuming LTE.
Importantly, the hydrodynamic models predict 
the velocity field $\bm{v}$, which is used by \mtd~to find
the Doppler shifts 
introduced by macroscopic gas flows at every gridpoint. 
The species being modelled is 
assumed to be a trace element
having no feedback on the background model atmosphere.
This is a good approximation for oxygen
which does not
contribute significantly to the continuous opacity,
nor to the number of free electrons.

The system of equations is solved
using the Multi-level Approximate Lambda Iteration (MALI)
preconditioning method 
of \citet{1991A&amp;A...245..171R,1992A&amp;A...262..209R}
\citep[see also][]{2001ApJ...557..389U},
in which, expressing the solution to \eqn{transportequation} 
as $I_{\nu}=\Psi_{\nu}\left[\eta_{\nu}\right]$,
the $\Psi_{\nu}$ operator 
is preconditioned using an approximate operator $\Psi^{*}_{\nu}$,
\phantomsection\begin{IEEEeqnarray}{rCl}
\label{preconditioning}
	I_{\nu} &=& \Psi^{*}_{\nu}\left[\eta_{\nu}\right] + \left(\Psi_{\nu}-\Psi^{*}_{\nu}\right)\left[\eta^{\dag}_{\nu}\right] .
\end{IEEEeqnarray}
Here $\eta_{\nu}$ is the emissivity and
$\eta^{\dag}_{\nu}$ is the emissivity determined
using the populations of the previous iteration.
\mtd~uses the diagonal of the $\Psi_{\nu}$ operator 
as the approximate operator.

The formal solution 
(the evaluation of \eqn{preconditioning})
is performed on short characteristics
\citep[the integral form of the radiative transfer equation
is solved analytically piecewise
along the ray;][]{1987JQSRT..38..325O}
using cubic-convolutuion interpolation of 
the upwind and downwind quantities,
and cubic Hermite spline interpolation 
for the source function
\citep{2003ASPC..288....3A,2003ASPC..288..419F,2013A&amp;A...549A.126I}.
At the lower boundary, the upwards intensity field
is taken to be that 
of a blackbody at the local material temperature,
while at the upper boundary it is assumed
that there is no incident radiation. 
Complete redistribution is assumed in the lines
and continua of the non-LTE atom,
and isotropic coherent scattering in all background processes.
The angle-averaged radiation field is calculated 
for 24 different directions using
\markcorr{Carlson's quadrature set 
A4 \citep{carlson1963numerical,1999A&amp;A...348..233B}.}

\markcorr{Acceleration of convergence techniques are important
both for achieving acceptable convergence rates and for avoiding
stabilisation of the 
solution \citep[e.g.][compare Fig.~3 and Fig.~4.]{1991ASIC..341....9A} 
The algorithm of \citet[][Appendix]{1990MWRv..118.1551K} was adopted,
which minimizes the global residual in the populations
with respect to a set of conjugate vectors 
between iterations.
We note briefly that their method is analogous to the
generalized conjugate residual algorithm 
\texttt{gcr}($m$), with restart parameter $m=2$,
that is presented in \citet[][]{saad2003iterative}
in the context of solving linear systems of equations.}

\markcorr{Once the populations have converged, 
a final formal solution is performed
to find the emergent specific intensities.
For this final formal solution,
interpolation of intensities is avoided
in order to reduce the numerical diffusion in the solution
\citep[long-characteristics; e.g.][]{2003ASPC..288....3A}.
This is achieved through
horizontal interpolation of the
extinction coefficients and source functions
onto the rays,
for every layer of the model atmosphere and every oblique ray.
The emergent intensity can then be found immediately
by direct integration.}
The final emergent intensities $\bar{I}_{\nu}$
are computed by averaging $I_{\nu}$ 
over the horizontal and temporal dimensions
using equidistant trapezoidal integration.
The emergent specific continuum intensity $\bar{I}^{\text{c}}_{\nu}$
are obtained in a similar fashion,
at each frequency point, but with 
zero contribution to the extinction and source function
from lines. 

The emergent flux 
(i.e.~the disk-integrated intensity)
\phantomsection\begin{IEEEeqnarray}{rCl}
\label{emergentflux}
	\bar{F}_{\nu} &=& \int_{0}^{2\uppi} \int_{0}^{1} \bar{I}_{\nu}\,\mu\,\ud\mu\,\ud\phi\, ,
\end{IEEEeqnarray}
and the emergent continuum flux,
\phantomsection\begin{IEEEeqnarray}{rCl}
\label{emergentcontinuumflux}
	\bar{F}^{\text{c}}_{\nu} &=& \int_{0}^{2\uppi} \int_{0}^{1} \bar{I}^{\text{c}}_{\nu}\,\mu\,\ud\mu\,\ud\phi\, ,
\end{IEEEeqnarray}
are calculated using
$n_{\mu}$-point Lobatto quadrature 
for the integral over $\mu$ 
on the interval [0,1] \citep[e.g.][]{1956itna.book.....H},
and equidistant $n_{\phi}$-point trapezoidal integration 
for the integral over $\phi$.
We used $n_{\mu}=5$ and, 
for $\mu\neq0$ and $\mu\neq1$,
we used $n_{\phi}=4$.
When $\mu=0$, the integrand is
identically nought and the formal
solution is not performed,
whilst when $\mu=1$ the ray is vertical and
only a single formal solution is required.
We note briefly that,
because of these properties of
the integrand, for a given number of formal solutions
Lobatto quadrature 
is of higher nominal accuracy than 
the closely related Gaussian quadratures
and Radau quadratures
\citep[e.g.][]{1956itna.book.....H}.

\subsection{Background opacities}
\label{methodbackground}

Continuous background opacities were calculated 
at runtime everywhere in the 3D stellar atmosphere 
using the Uppsala opacity package 
\citep[][and subsequent updates]{1975A&amp;A....42..407G}.
These opacities were calculated 
under the assumption of LTE.
Oxygen continua that were also present in
the model atom (discussed below) were
not included at this point. 

Background line opacities in the UV
were pre-computed,
in pure absorption and assuming LTE,
on a grid of gas temperatures and pressures
and with an assumed equation of state
for the appropriate chemical compositions of the models
\citep{2005A&amp;A...442..643C,2012MNRAS.427...27B}.
The line opacities were interpolated
onto the model atmosphere by \mtd.
In particular, 
the $121.5\,\mathrm{nm}$ Lyman-$\alpha$ transition
was included as a background opacity 
in this way; we found this line
to have a dramatic effect on
the statistical equilibrium at low metallicities (see \sect{resultslyman}).

\subsection{Model atom}
\label{methodatom}

\begin{figure}
\begin{center}
\includegraphics[scale=0.31]{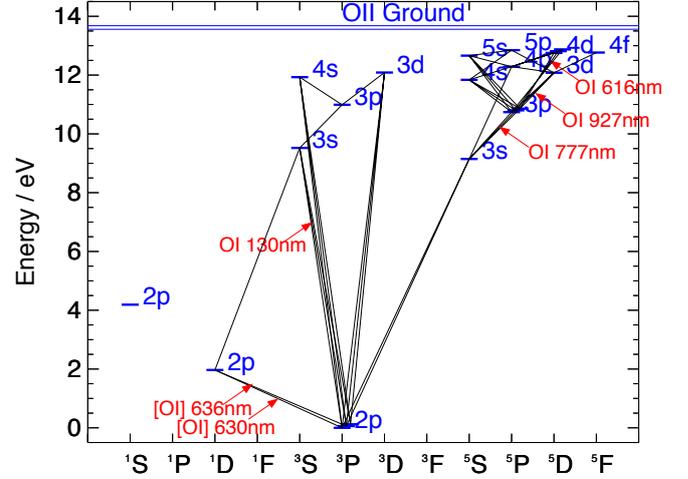}
\caption{Grotrian diagram of the 22 levels of oxygen
and the 43 radiative transitions
included in the model atom.
Fine structure has been magnified.
The ground state of OII is also included
in the model atom,
along with 22 bound-free 
radiative transitions 
(not depicted).}
\label{grotrian}
\end{center}
\end{figure}

\begin{table*}
\begin{center}
\caption{Collisional transitions included in the model atom. 
Charge transfer was only included for the 
oxygen ground state.
Collisional rate coefficients $q$ are obtained from
collisional cross-sections $\sigma$ 
via the relation $q=<\sigma v>$;
here the brackets indicate averaging over 
the Maxwell-Boltzmann distribution for speeds $v$.
Assuming that the particle speeds
have Maxwell-Boltzmann distributions,
the forward and reverse rate
coefficients are related 
by the Boltzmann factor \citep[e.g.][]{2003rtsa.book.....R}.}
\label{collisionstable}
\begin{tabular}{c r c l c c }
\hline
Transition & \multicolumn{3}{c}{Equation} & Reference & Comment   \\
\hline
\hline
Electron excitation & $\mathrm{O} + \mathrm{e^{-}}$ &$\longleftrightarrow$&  $\mathrm{O^{*}} + \mathrm{e^{-}}$ & {\citet{2007A&amp;A...462..781B}} & Ab initio; quantum mechanical  \\
\hline
Electron ionisation & $\mathrm{O} + \mathrm{e^{-}}$ &$\longleftrightarrow$&  $\mathrm{O^{+}} + 2\mathrm{e^{-}}$ & {\citet{1973asqu.book_rate.....A}} & Empirical \\
\hline
Neutral hydrogen excitation & $\mathrm{O} + \mathrm{H}$ &$\longleftrightarrow$&  $\mathrm{O^{*}} + \mathrm{H}$ & {\citet{1993PhST...47..186L}} & Semi-empirical; semi-classical  \\
\hline
Neutral hydrogen ionisation & $\mathrm{O} + \mathrm{H}$ &$\longleftrightarrow$&  $\mathrm{O^{+}} + \mathrm{H} + \mathrm{e^{-}}$ & {\citet{1993PhST...47..186L}}  & Semi-empirical; semi-classical \\
\hline
Proton charge transfer &$\mathrm{O} + \mathrm{H^{+}}$ &$\longleftrightarrow$&  $\mathrm{O^{+}} + \mathrm{H}$ & {\cite{1999A&amp;AS..140..225S}} & Mixed \\
\hline
\end{tabular}
\end{center}
\end{table*}

The model oxygen atom that was
used in this work is based on those used by  
\citet{1993ApJ...402..344C},
\citet{1993A&amp;A...275..269K},
and \citet{2009A&amp;A...500.1221F}.
We show in \fig{grotrian} the
radiative transitions connecting
the 23 levels (22 excited levels plus 1 ionised level)
included in the model.
The model was updated to include  
energies and Einstein coefficients
\markcorr{from 
NIST\footnote{\url{http://www.nist.gov/pml/data/asd.cfm}} \citep{NIST_ASD}},
natural line broadening coefficients
\markcorr{from 
VALD3\footnote{\url{http://vald.astro.uu.se/~vald/php/vald.php}}
\citep{1995A&amp;AS..112..525P,2008JPhCS.130a2011H}},
and collisional line broadening coefficients from
\citet{1998PASA...15..336B}.
Photoionisation cross-sections were the same
as those used in the papers cited above;
i.e.~taken from 
the Opacity Project \citep{1993A&amp;A...275L...5C},
or calculated semi-empirically \citep{1971JPhB....4.1670P}.
\markcorr{Oscillator strengths for the \lnf{630}~and \lnf{636}~lines
were taken from \citep{2000MNRAS.312..813S},
and atomic data for the Ni blend in the former line
were taken from \cite{2003ApJ...584L.107J}.}

We show in \tab{collisionstable}
the collisional transitions
included in the model
and their sources.
The collisional rates between
fine structure levels were set to very
large values, which ensured that the 
populations within the fine structure
were distributed according to their
statistical weights \citep[e.g.][]{1993A&amp;A...275..269K}.
Following \citet{1993PhST...47..186L},
a correction factor $\sh$ to the original formula
of \citet{1968ZPhy..211..404D,1969ZPhy..225..483D} 
was adopted to describe 
neutral hydrogen collisions.
Levels with radiatively weak transitions were coupled 
using an effective oscillator strength
$f_{\mathrm{min}}=10^{-3}$.
Following \citet{2009A&amp;A...508.1403P},
and using their
observational data from the Swedish 1-m Solar Telescope
\citep{2003SPIE.4853..341S},
the centre-to-limb variation in the Sun
of the \lna{777}~lines were used
to calibrate $\sh$.
We obtained $\sh\approx1$
and inferred a solar oxygen abundance $\lgeps^{\odot}\approx8.7$.
A detailed description of the fitting procedure 
will be presented in a forthcoming paper.

\subsection{Model atmospheres}
\label{methodmodelatmospheres}

Line formation calculations
were performed across a grid of 3D hydrodynamic 
\stagger~model atmospheres \citep{2011JPhCS.328a2003C,
2013A&amp;A...557A..26M}.
Each model in the grid is specified
by a nominal effective temperature $\teff$,
surface gravity $\lgg$\footnote{Here
and henceforth, $\log\left(...\right)\equiv\log_{10}\left(...\right)$.},
and \feh\footnote{The logarithmic abundance
of an arbitrary element A is defined with respect to hydrogen:
$\log\epsilon_{\text{A}}=\log\frac{N_{\text{A}}}{N_{\text{H}}}+12$.
The abundance ratio of elements A and B is given by:
$[\text{A}/\text{B}]=(\log\epsilon_{\text{A}}-\log\epsilon_{\text{A}}^{\odot})-(\log\epsilon_{\text{B}}-\log\epsilon_{\text{B}}^{\odot})$,
where $\odot$ denotes the solar value.}.
Solar scaled abundances \citep{2009ARA&amp;A..47..481A}
were adopted, with $\alpha$-enhancement 
$\left[\alpha/\text{Fe}\right]=0.4$ for models
with $\feh\leq-1.0$.
Since oxygen was treated as a trace element,
the oxygen abundance was varied independently.
Calculations were performed on four
snapshots for each 3D model,
equally spaced across
a time sequence spanning approximately two
convective turnover times \citep{2013A&amp;A...557A..26M}.

The original models have Cartesian geometry 
with $xyz$-grid size $240\times240\times240$.
These were modified in several steps before
performing the detailed line formation calculations.
First, every other gridpoint in the horizontal dimensions
was dropped, reducing the number of
gridpoints by a factor of four.
Second, unphysical ghost layers (5 layers 
on both the upper and the lower boundaries)
were removed as well as
additional layers
such that the top-most layer
satisfied $\mathrm{Max}\left(\lgt\right)\lesssim-5$
and that the bottom-most layer
satisfied $\mathrm{Min}\left(\lgt\right)\gtrsim3$,
where $\lgt$ is the 
vertical optical depth at wavelength $\lambda=500\,\text{nm}$.
Finally, $T$, $\log{\rho}$, $\log{n_{\text{e}}}$ and $\bm{v}$ were 
interpolated using cubic splines onto a new vertical grid.
This new vertical grid was constructed by demanding that
the horizontal-mean step in $\lgt$ was roughly constant
between layers.
The final grid size was fixed 
at $120\times120\times110$
(where the last dimension represents the vertical). 

Line formation calculations were also performed
on 1D plane-parallel hydrostatic
\atmo~\citep[][Appendix A]{2013A&amp;A...557A..26M} model atmospheres.
These models were computed using an equation of state
and a treatment of opacities and radiative transfer fully 
consistent with the \stagger~models.
They have 110 depth points and span
$-5\leq\lgr\leq3$,
where $\lgr$ is the logarithmic
vertical Rosseland mean optical depth.
The models use mixing length theory 
\markcorr{\citep[][]{1958ZA.....46..108B}}
to account for convective energy transport
in an approximate way.
\markcorr{The adopted mixing length parameter 
was $\alpha_{\mathrm{MLT}}=1.5$
(in units of pressure scale heights).
This parameter affects the atmospheric structure
in the deepest layers, below where the relevant oxygen lines form;
hence, the results are insensitive to the precise value of this parameter.}
The original opacity sampling data
\markcorr{were} computed assuming
a microturbulence parameter
of $1.0\,\kms$;
these opacities are sorted into 12 bins
in exactly the same way as for the
corresponding 3D \stagger~simulations.
Apart from a different choice of angle
quadrature for the mean intensity
(10 directions, based on 5-point Gaussian quadrature
on the interval $\mu\in[0,1]$)
and the use of another microturbulence parameter $\xi$
(discussed in \sect{methodgrids}),
the 1D line formation calculations
proceeded in the same way as the 3D calculations
using \mtd.

\subsection{Equivalent widths, abundance corrections
and abundance errors}
\label{methodgrids}

\begin{table}
\begin{center}
\caption{Nominal effective temperatures,
and corresponding surface gravities, on the nodes on the grid.}
\label{gridtable1}
\begin{tabular}{c c c c c c}
\hline
Nominal $\teff/\mathrm{K}$ & \multicolumn{5}{c}{$\lggu$} \\
\hline
\hline
5000 & 3.0 & 3.5 & 4.0 & 4.5 & 5.0 \\
\hline
5500 & 3.0 & 3.5 & 4.0 & 4.5 & 5.0 \\
\hline
6000 &  & 3.5 & 4.0 & 4.5 &  \\
\hline
6500 & &  & 4.0 & 4.5 &  \\
\hline
\end{tabular}
\end{center}
\end{table}

\begin{table}
\begin{center}
\caption{Background chemical compositions,
and corresponding \markcorr{non-LTE} oxygen abundances, on the nodes
of the grid. \markcorr{LTE runs were performed
for all oxygen abundances $5.7\leq\lgeps\leq10.7$
in regular intervals of 0.5 dex.}}
\label{gridtable2}
\begin{tabular}{c c c c c c c c c c}
\hline
$\feh$ & \multicolumn{9}{c}{$\lgeps$} \\
\hline
\hline
-3.0 & 5.7 & 6.2 & 6.7 & 7.2 & 7.7 & & & &  \\
\hline
-2.0 & 5.7 & 6.2 & 6.7 & 7.2 & 7.7 & 8.2 & 8.7 & &  \\
\hline
-1.0 &  &  & 6.7 & 7.2 & 7.7 & 8.2 & 8.7 & 9.2 &  \\
\hline
0.0 &  &  &  &  & 7.7 & 8.2 & 8.7 & 9.2 & 9.7 \\
\hline
\end{tabular}
\end{center}
\end{table}

To obtain equivalent widths,
the non-LTE/LTE ratios of the
averaged emergent intensities from \mtd~were found
and applied to
the 3D LTE averaged emergent intensities calculated 
using \scate~\markcorr{\citep{2010A&amp;A...517A..49H}}
over at least 50 temporal snapshots across the entire time sequence.
The intensities were averaged
not just over the spatial and temporal dimensions,
but also over the the azimuthal angle,
using equidistant trapezoidal integration.
The flux was then computed using \eqn{emergentflux}.
This procedure was repeated
using the continuum intensity
(\eqn{emergentcontinuumflux});
thereby the normalized flux could be obtained.
\markcorr{It is better to use intensity ratios at this stage
instead of flux ratios; if the systematic errors associated with using
just four snapshots have any $\mu$~dependence, the error in the flux ratio
will be larger than the error in the intensity ratios.}
\markcorr{The equivalent widths were found by direct integration.}

We define abundance corrections
as the abundance differences compared to 1D LTE.
For example, the 3D non-LTE abundance correction,
\phantomsection\begin{IEEEeqnarray}{rCl}
\label{abundancecorrection}
    \corr{3N}{1L}\left(\lgeps^{\mathrm{1D,LTE}}\right)&=&\lgeps^{\mathrm{3D,NLTE}}-\lgeps^{\mathrm{1D,LTE}},
\end{IEEEeqnarray}
is determined by demanding that
the equivalent width of the line flux
obtained in LTE from the 1D model
with a given oxygen abundance ($\lgeps^{\mathrm{1D,LTE}}$)
matches that obtained from the 3D non-LTE grid 
with some oxygen abundance ($\lgeps^{\mathrm{3D,NLTE}}$).
It is a function of the 1D LTE abundance
and the stellar parameters.
In contrast, we define abundance errors
as the abundance differences compared to 3D non-LTE.
For example, the 1D LTE abundance error,
\phantomsection\begin{IEEEeqnarray}{rCl}
\label{abundanceerrors}
    \corr{1L}{3N}\left(\lgeps^{\mathrm{3D,NLTE}}\right)&=&\lgeps^{\mathrm{1D,LTE}}-\lgeps^{\mathrm{3D,NLTE}},
\end{IEEEeqnarray}
is a function of the 3D non-LTE abundance
and the stellar parameters. 
Abundance errors are more illuminating 
in the context of this work and
are discussed in \sect{resultsabundanceerrors}.
However, abundance corrections 
to 1D LTE based \markcorr{results} are more immediately applicable 
to surveys of FGK-type stars;
\markcorr{such corrections have already been illustrated in a previous 
publication \citep{2015MNRAS.454L..11A},
and are tabulated in this work in \sect{resultsgrids}.} 

The extent of the grids is summarised
in \tab{gridtable1} and \tab{gridtable2}.
The actual (time-averaged) 
effective temperatures of the 3D hydrodynamic
simulations are not input parameters,
and for a given simulation
is slightly different to its nominal effective temperature
\citep{2013A&amp;A...557A..26M}.
The effective temperatures of the 1D models
were set to match those of their 3D counterparts.
With the 1D models, line formation calculations 
were performed using microturbulence parameters
$\xi=$ 0.5, 1.0, 1.5 and 2.0 $\kms$.
This parameter is necessary to account
for the line broadening by convective velocity fields,
which are not predicted in the 1D models.

\section{Results} 
\label{results}

\subsection{Non-LTE effects}
\label{resultsnonlte}

\begin{figure*}
\begin{center}
\includegraphics[scale=0.66]{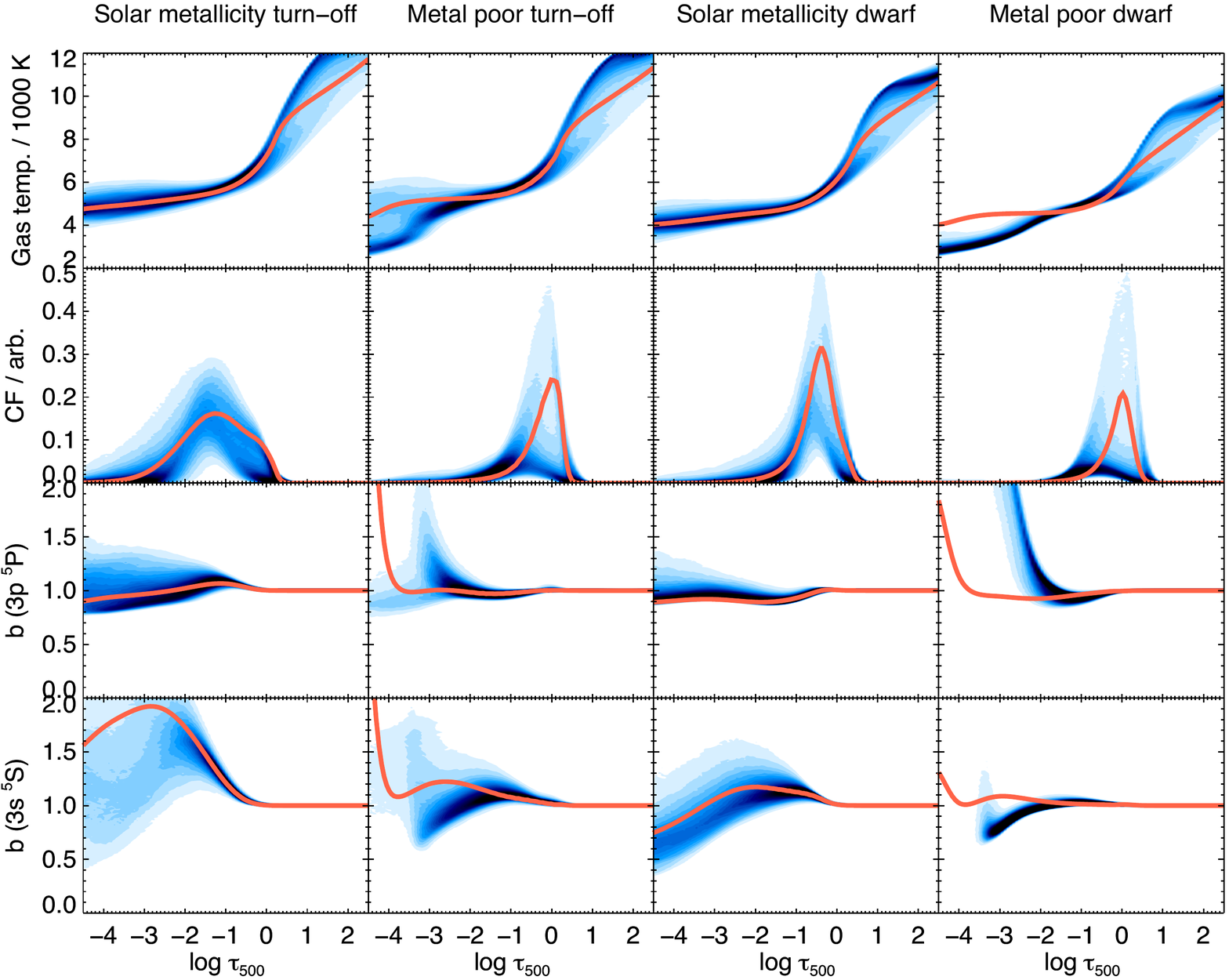}
\caption{The 
departures from LTE of the \lna{777}~lines
at different depths
in a turn-off star ($\teff\approx6500\,\mathrm{K}$, $\lggu=4.0$)
and in a dwarf ($\teff\approx5500\,\mathrm{K}$, $\lggu=4.0$) 
with $\feh=0.0$, $\lgeps=8.7$ (solar metallicity)
and $\feh=-3.0$, $\lgeps=6.2$ (metal poor).
The quantities from top to bottom are
gas temperature, 
\markcorr{non-LTE frequency-integrated 
contribution function for the absolute 
line flux depression (in arbitrary units)},
departure coefficient of the \tripletup~level 
(the upper level of the \lna{777}~lines),
and departure coefficient of the \tripletlo~level 
(the lower level of the \lna{777}~lines).
\markcorr{These quantities were evaluated on surfaces
of constant $\lgt$.
Corresponding quantities from the 1D models
are overplotted (red solid lines).}}
\label{tcbb}
\end{center}
\end{figure*}

\begin{figure*}
\begin{center}
\includegraphics[scale=0.66]{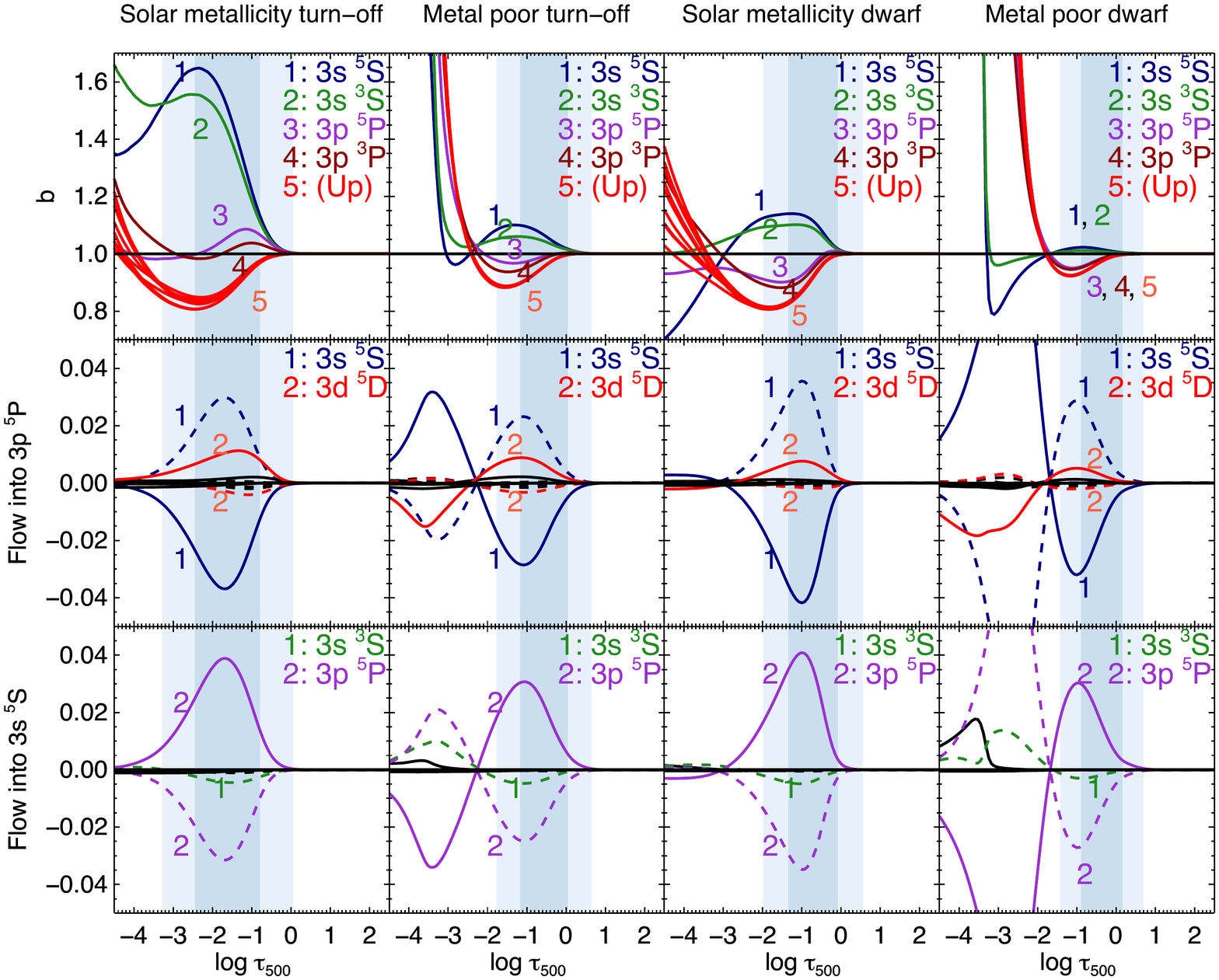}
\caption{The statistical equilibrium state
at different depths
in a turn-off star ($\teff\approx6500\,\mathrm{K}$, $\lggu=4.0$) 
and in a dwarf star ($\teff\approx5500\,\mathrm{K}$, $\lggu=4.0$) 
with $\feh=0.0$, $\lgeps=8.7$ (solar metallicity)
and $\feh=-3.0$, $\lgeps=6.2$ (metal poor).
The top row shows the departure coefficients of each level;
the 2p levels in the spin-0 and spin-1 systems
as well as the OII ground level do not experience 
departures from their LTE populations
and are not shown here.
`Up' refers to the 4s $^{5}$S level, and higher energy levels.
The second and third rows show
the normalized net flow into the \tripletup~level
(the upper level of the \lna{777}~lines)
and the normalized net flow into into the \tripletlo~level
(the lower level of the \lna{777}~lines).
Collisional transitions are indicated as dashed lines
while radiative transitions are indicated as solid lines.
Plotted here are median values
on surfaces of equal $\lgt$.
The shaded areas correspond to the
mean formation depth 
$\pm$ one standard deviation,
and $\pm$ two standard deviations.
\markcorr{The mean formation depth is defined here as the
expectation of $\lgt$ with respect to the 
contribution function $C$ in the 3D volume $V$:
$\expect{\lgt}=\frac{\int{\lgt\,C\,\ud V}}{\int{C\,\ud V}}$.
The standard deviation is given by
$\sqrt{\expect{\lgt^{2}}-\expect{\lgt}^{2}}$.}
}
\label{bdd}
\end{center}
\end{figure*}

The departure coefficients
i.e.~the ratios of the non-LTE 
statistical equilibrium populations
to the LTE populations,
\phantomsection\begin{IEEEeqnarray}{rCl}
\label{departurecoefficients}
    b_{i} &=& \frac{n_{i}}{n^{*}_{i}}\, ,
\end{IEEEeqnarray}
are quantitative measures of the departures from LTE.
In the Wien regime, the line source function
approximately scales with $\frac{b_{u}}{b_{l}}$
and the line extinction 
approximately scales with $b_{l}$
\citep[e.g.][]{2003rtsa.book.....R}.
Used together with line contribution functions
\citep[e.g.][]{1986A&amp;A...163..135M,
1996MNRAS.278..337A,2015MNRAS.452.1612A},
which can be used to identify the line forming regions,
the departure coefficients are 
illuminating and intuitive diagnostics of
non-LTE line formation.

We plot in \fig{tcbb} the probability 
distributions of the gas temperature,
\markcorr{the non-LTE frequency-integrated 
contribution function for the absolute 
line flux depression} \citep{2015MNRAS.452.1612A}, 
and the departure coefficients of the \tripletup~level
and the \tripletlo~level
(i.e.~the upper and lower levels of the \lna{777} lines)
across the atmosphere.
We show this for four representative theoretical stars:
a turn-off star and a dwarf,
each with $\feh=0.0$ and $\feh=-3.0$.

At large depths, densities are large enough to
guarantee that collisional rates dominate over
all radiative rates, such that the populations of
all levels do not depart significantly from their LTE 
values. Higher up in the atmosphere,
the \lna{777}~lines form
(second row of \fig{tcbb}), and the
\tripletup~level and the \tripletlo~level
populations show departures from their LTE 
values. This reflects the fact
that \markcorr{photon losses in} the lines themselves drive
the non-LTE effects in these levels
\markcorr{\citep[e.g.][]{2005ARA&amp;A..43..481A}},
as discussed below.

For the stars shown, as well as more generally across this 
range of stellar parameters,
the \tripletlo~level develops
a significant overpopulation with respect to LTE
in the line-forming regions.
As the atmosphere becomes less homogeneous, 
the spread of departure coefficients
on surfaces of equal optical depth  
becomes greater with height.
The behaviour of the \tripletup~level 
is more complicated. 
Generally, in the line forming regions,
the \tripletup~level stays close
to unity or develops a slight
underpopulation with respect to LTE.
The exception is in stars with high
effective temperatures,
low surface gravities and 
higher oxygen abundances
(e.g.~the solar metallicity turn-off star 
in the first column of \fig{tcbb}),
where the line forming regions are quite extended.
In such stars, around the continuum forming layers
the \tripletup~level develops a slight overpopulation
with respect to LTE, while higher up, it typically 
develops an underpopulation with respect to LTE.

Across the 3D \stagger~grid, the overpopulations of
the \tripletlo~level tend to be larger
than the departures from LTE in the \tripletup~level.
Therefore in general, the line opacity is increased,
while the line source function is decreased,
compared to the LTE case.
Both of these effects increase the line 
contribution function and 
hence the strength of the \lna{777}~lines with respect to LTE.

To understand the origin of these various effects,
the behaviours of all of the energy levels
need to be considered.
We show on the first row of \fig{bdd} 
the median departure coefficients for all levels
across the same set of stars.
The OI ground level (\oground)
does not deviate significantly from its 
LTE population throughout the atmosphere,
as expected given that oxygen remains overwhelmingly 
neutral.
The efficient charge transfer reaction ensures that
the OII ground level also maintains its LTE population.
Similarly, the first two excited levels \term{2}{p}{1}{S}~and 
\term{2}{p}{1}{D}~do not deviate
significantly from their LTE populations,
due to efficient collisional coupling with the ground level
and the lack of any strong radiative couplings.
The \tripletup~level (the upper level of the \lna{777}~lines)
and the \term{3}{p}{3}{P}~level (its spin-one analogue)
behave similarly to each other,
as do the \tripletlo~level (the lower level of the \lna{777}~lines)
and the \term{3}{s}{3}{S}~level (its spin-one analogue),
as a result of efficient intersystem coupling
by electron collisions
\citep{2007A&amp;A...462..781B}.
The highly excited upper levels (\term{4}{s}{5}{S}~and above)
are efficiently coupled by neutral hydrogen collisions,
thus their departure coefficients
behave similarly to each other;
they tend to be underpopulated
in the line forming regions,
for reasons discussed below.

The departure coefficients can in turn 
be understood by considering the net flow
into level $j$ from level $i$:
\phantomsection\begin{IEEEeqnarray}{rCl}
\label{rhoequation}
    \rho^{T}_{i\,j} &=& n_{i}\,T_{i\,j} - n_{j}\,T_{j\,i},
\end{IEEEeqnarray}
with $T_{i\,j}=R_{i\,j}$ or $T_{i\,j}=C_{i\,j}$ for
radiative and collisional transitions, respectively.
In detailed balance, $\rho^{T}_{i\,j}$ 
is zero for all values of $i$ and $j$.

We show on the second and third rows of \fig{bdd} 
the median net flow into the \tripletup~level 
and into the \tripletlo~level
for radiative and collisional transitions,
normalized using the total flow into 
those respective levels.
In the line forming regions,
the net flow is largely dictated by 
the radiative coupling between these levels
(i.e.~the \lna{777}~lines),
balanced by the collisional coupling between these levels.
The \lna{777}~lines act to shift the equilibrium away from 
LTE to one where the \tripletup~level is underpopulated
and the \tripletlo~level is overpopulated.

\markcorr{The mechanism for this is photon losses.
Photons that are absorbed by the lines are either destroyed
via collisional de-excitation (true absorption),
or are scattered, via spontaneous or by stimulated photoemission
\citep[e.g.][]{2003rtsa.book.....R}.
Scattered photons can propagate large
distances through the atmosphere before being absorbed again
(and possibly scattered again), or can escape from the
atmosphere altogether.
Consequently, there are fewer photoexcitations
than there would be given a thermal radiation field,
so that the upper level becomes underpopulated
and the lower level becomes overpopulated, relative to LTE.
This mechanism is particularly effective for the \lna{777}~lines,
which have extended Lorentzian wings.
Photons absorbed at the line-centre, where the extinction probability is large,
can be scattered into the wings (since complete redistribution is assumed);
the extinction probability in the wings is small,
so such photons are very likely to propagate large distances and escape.
For more details on this mechanism 
see e.g.~chapter 14 of \citet{2014tsa..book.....H}.}

\markcorr{Other radiative transitions
are also important to the behaviours of the
\tripletlo~and \tripletup levels.
In particular, there}
is a cascade from the collisionally-coupled 
highly excited upper levels 
down to the \tripletup~level 
(and hence to the \tripletlo~level,
via the \lna{777}~lines)
via mainly the \term{3}{d}{5}{D} level, or the \lna{927}~line.
As pointed out by \citet{1993A&amp;A...275..269K},
\markcorr{photon losses in the}
radiative transitions originating in these levels 
act to shift the equilibrium away from 
LTE to one where the upper levels are underpopulated
and where the lower levels
are overpopulated,
analogous to the way 
the \lna{777}~lines act on the \tripletup~level
and the \tripletlo~level.
There is thus a flow into the \tripletup~level 
which competes with the
underpopulating effect of the \lna{777}~lines.
This competing effect complicates
the behaviour of the \tripletup~level
and is what drives the overpopulation
seen at $\lgt\approx-1.0$ in the
solar metallicity turn-off star in \fig{tcbb}.
In contrast, the behaviour of the \tripletlo~level
is simpler to understand. The \lna{777}~lines
induce a flow into the level, which is metastable
and can develop a relatively large overpopulation.

The relative size of these various effects are sensitive to  
the stellar parameters.
Larger effective temperatures, smaller 
effective gravities and larger oxygen abundances
tend to make the \lna{777}~lines stronger.
which tends to shift the system further away from LTE.
They also make the higher excitation lines
(such as the \lna{927}~line) stronger.
This increases the cascade effect 
described above, and is why the overpopulation
in the \tripletup~levels
is most prominent in the more metal rich
turn-off stars.

The preceding discussion focused on the
line forming regions of the \lna{777}~lines.
At greater heights, the lines
do not form efficiently because the populations
of the high excitation levels are negligible
in these low temperature regions.
A discussion of the limiting
behaviour of the departure coefficients
and of $\rho^{T}_{i\,j}$ in these regions
is therefore of lesser interest.

\subsection{3D atmospheric inhomogeneities}
\label{resultsinhomogeneities}

\begin{figure*}
\begin{center}
\includegraphics[scale=0.66]{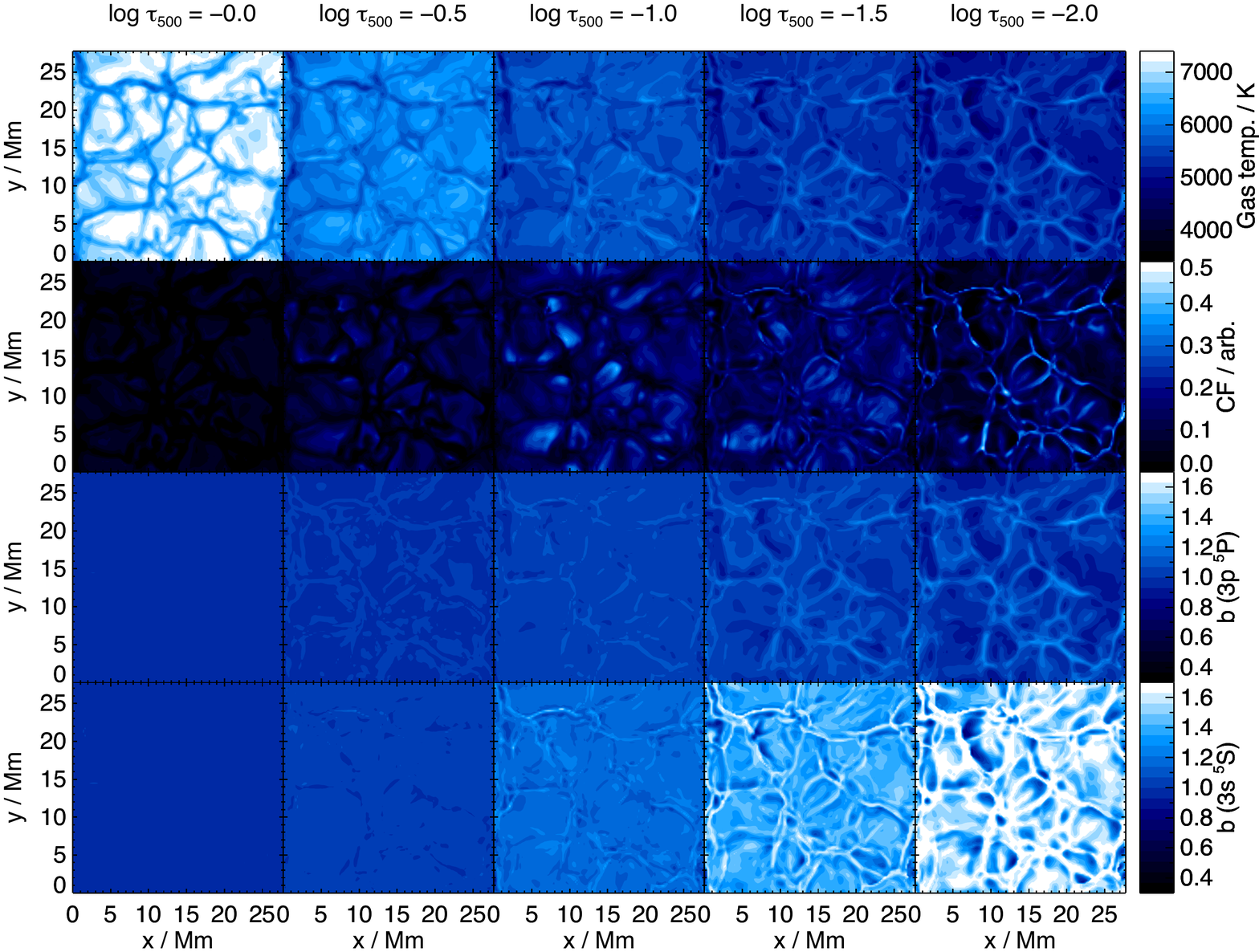}
\caption{Surfaces of equal vertical optical depth $\lgt$
in a snapshot of a 
turn-off star ($\teff\approx6500\,\mathrm{K}$, $\lggu=4.0$) 
with $\feh=0.0$,
and oxygen abundance $\lgeps=8.7$.
The quantities from top to bottom are
gas temperature, 
\markcorr{non-LTE frequency-integrated 
contribution function for the absolute 
line flux depression (in arbitrary units)},
departure coefficient of the \tripletup~level
(the upper level of the \lna{777}~lines),
and departure coefficient of the \tripletlo~level
(the lower level of the \lna{777}~lines).
The mean formation depth ($\pm$ one standard deviation),
\markcorr{as defined in \fig{bdd}},
is $\expect{\lgt}\approx-1.66\left(\pm0.86\right)$.}
\label{map1}
\end{center}
\end{figure*}

\begin{figure*}
\begin{center}
\includegraphics[scale=0.66]{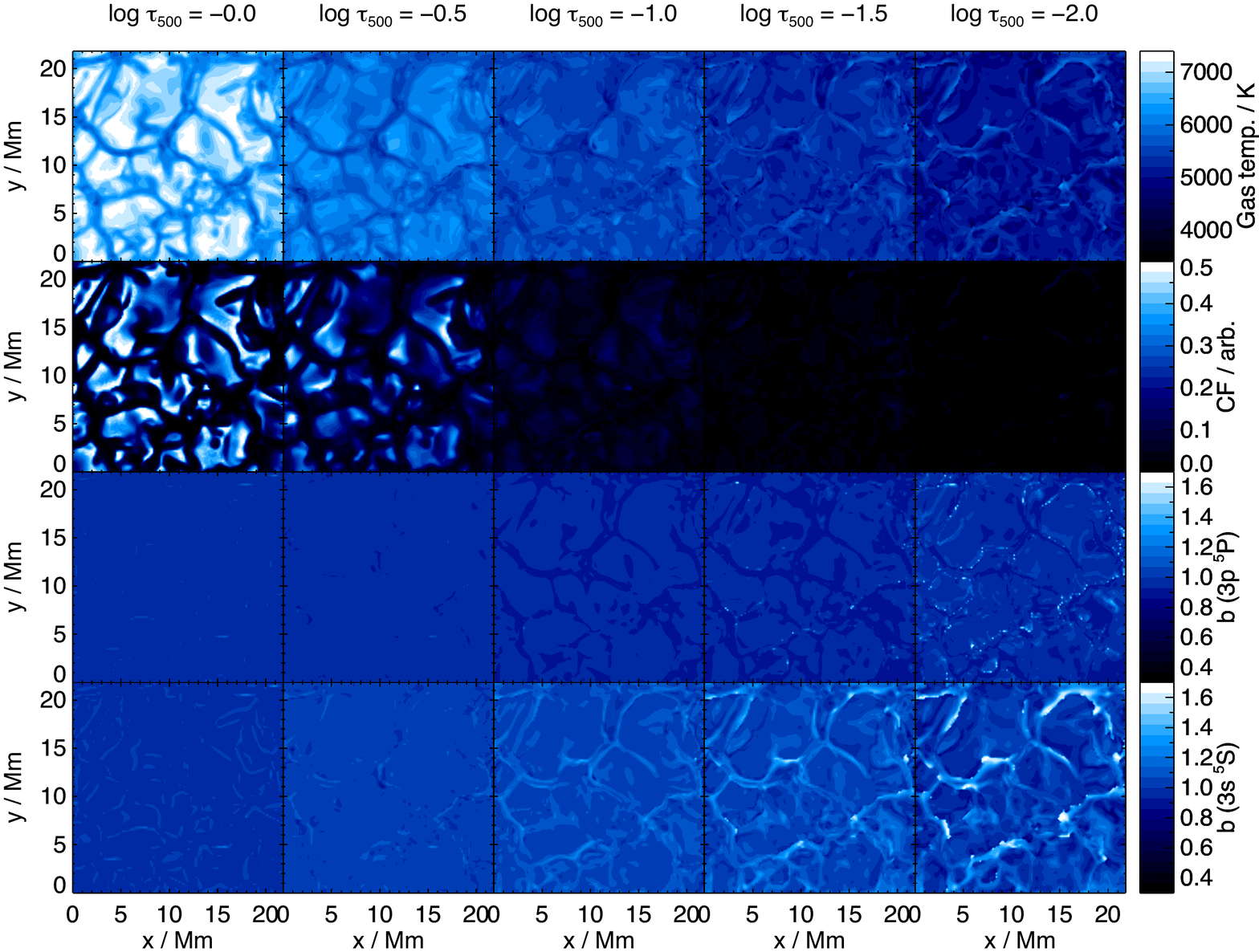}
\caption{Surfaces of equal vertical optical depth $\lgt$
in a snapshot of a 
turn-off star ($\teff\approx6500\,\mathrm{K}$, $\lggu=4.0$) 
with $\feh=-3.0$,
and oxygen abundance $\lgeps=6.2$.
The quantities from top to bottom are
gas temperature, 
\markcorr{non-LTE frequency-integrated 
contribution function for the absolute 
line flux depression (in arbitrary units)},
departure coefficient of the \tripletup~level
(the upper level of the \lna{777}~lines),
and departure coefficient of the \tripletlo~level
(the lower level of the \lna{777}~lines).
The mean formation depth ($\pm$ one standard deviation),
\markcorr{as defined in \fig{bdd}},
is $\expect{\lgt}\approx-0.55\left(\pm0.59\right)$.}
\label{map2}
\end{center}
\end{figure*}

We now consider the effects of
atmospheric inhomogeneities
on the statistical equilibrium.
We show in \fig{map1} 
the spatially-resolved gas temperature,
\markcorr{the non-LTE frequency-integrated 
contribution function for the absolute 
line flux depression}, 
and departure coefficients of the \tripletup~level
and the \tripletlo~level
(i.e.~the upper and lower levels of the \lna{777} lines)
on five maps of equal $\lgt$
in a single 3D snapshot of a solar metallicity turn-off star.
In \fig{map2} we show the same quantities,
in a snapshot of a metal poor turn-off star.

At the largest depths shown ($\lgt=0.0$,
corresponding roughly to the visible surface)
the granulation pattern which results
from convective turnover at the bottom of the photosphere
\citep[e.g.][]{2013A&amp;A...557A..26M}
is evident in the gas temperature maps. 
The bright granules occupying the largest area fraction
are composed of hot upflowing gas;
cooler downflowing gas make up the intergranular lanes.
At the smallest depths shown ($\lgt=-2.0$),
reverse granulation features
\citep[e.g.][]{2007A&amp;A...461.1163C}
are apparent in the gas temperature maps.
The upflowing gas cools 
efficiently by adiabatic expansion,
while radiative reheating and mechanical compression
cause the downflowing gas to
increase slightly in temperature
\citep{2013A&amp;A...560A...8M}.

Most of the emergent \lna{777}~line formation
(as indicated by the 
\markcorr{contribution function for the absolute line flux depression})
occurs at depths $\lgt\lesssim0.5$,
because photons from larger depths
are unlikely to escape through the atmosphere
(cf. the escape probability $\expo{-\tau}$). 
Around the optical surface
($-1.0\lesssim\lgt\lesssim0.0$),
the line formation is associated
with the granulation pattern,
while at greater heights
($\lgt\lesssim-1.5$)
in the solar metallicity example,
line formation is associated with 
the reverse granulation features.
(in the metal poor example (\fig{map2}),
line formation is inefficient 
on account of the low oxygen abundance.)
These associations are 
because the line opacity of 
the high-excitation \lna{777}~lines
is larger in regions of higher gas temperatures,
due to greater excitation out of the oxygen ground level.

At large depths ($\lgt\gtrsim0.5$),
the departure coefficients stay close to unity
because the large densities guarantee 
that collisional rates dominate over all radiative rates.
Around the optical surface
($-1.0\lesssim\lgt\lesssim0.0$),
the \tripletlo~level develops 
an overpopulation with respect to LTE
in both \markcorr{figures;}
in \markcorr{\fig{map1}} the \tripletup~level also 
becomes overpopulated.
At greater heights ($\lgt\lesssim-1.5$) \markcorr{in \fig{map1}}
the overpopulations in the levels
are associated with the reverse granulation features. 
\markcorr{This behaviour
can be understood from the preceding discussion of the 
non-LTE mechanism (\sect{resultsnonlte}).}
The departures from LTE
in the \tripletlo~and \tripletup~levels
are driven largely by \markcorr{photon losses in} the
high excitation \lna{777}~lines,
and are also affected by \markcorr{photon losses in} even 
higher excitation lines such as the \lna{927}~lines.
\markcorr{The general trend of increasing 
departure coefficient with height 
is due to photon losses becoming more severe
as the material becomes increasingly optically thin.
On top of that, photon losses are more severe
in regions of high temperature,
where the lines are stronger;
consequently there is a close association of the
departure coefficients with the reverse granulation features.}

The efficient line formation
of the \lna{777}~lines in the reverse granulation is
interesting because it drives
an overpopulation with respect to LTE in the
\tripletlo~level, which in turn
promotes line formation. This is 
an example of the coupling of
3D effects with non-LTE effects.

\subsection{Non-vertical radiative transfer}
\label{results15d}

\begin{figure}
\begin{center}
\includegraphics[scale=0.66]{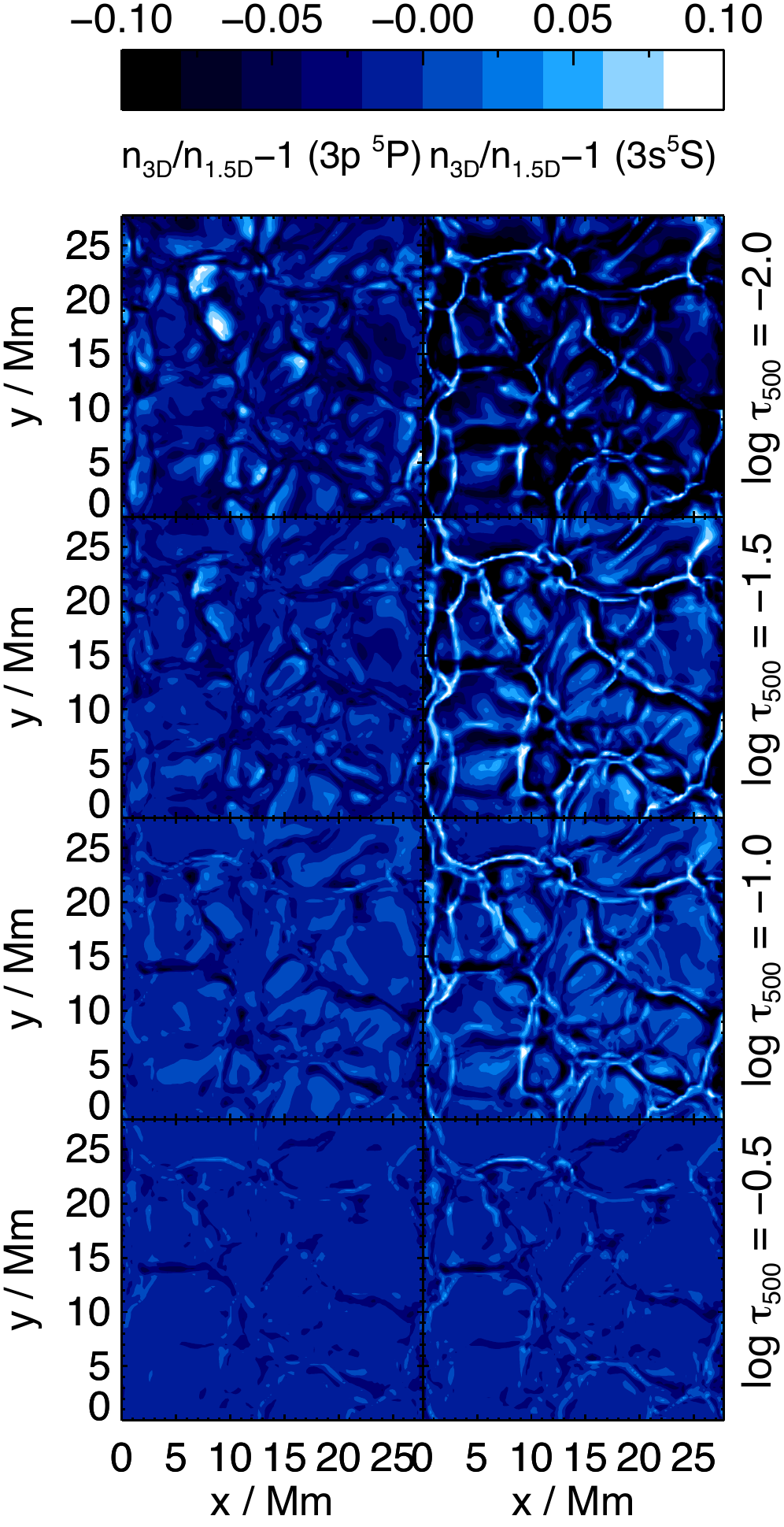}
\caption{Relative difference in the departure coefficients of the
\tripletup~level (left) and of the \tripletlo~level (right)
(i.e.~the upper and lower levels of the \lna{777}~lines,
respectively)
between the 1.5D non-LTE case (which neglects
non-vertical radiative transfer)
and the 3D non-LTE case,
on surfaces of equal vertical optical depth $\lgt$
in a snapshot of a 
turn-off star ($\teff\approx6500\,\mathrm{K}$, $\lggu=4.0$) 
with $\feh=0.0$,
and oxygen abundance $\lgeps=8.7$.}
\label{15dratio}
\end{center}
\end{figure}

\begin{figure*}
\begin{center}
\includegraphics[scale=0.66]{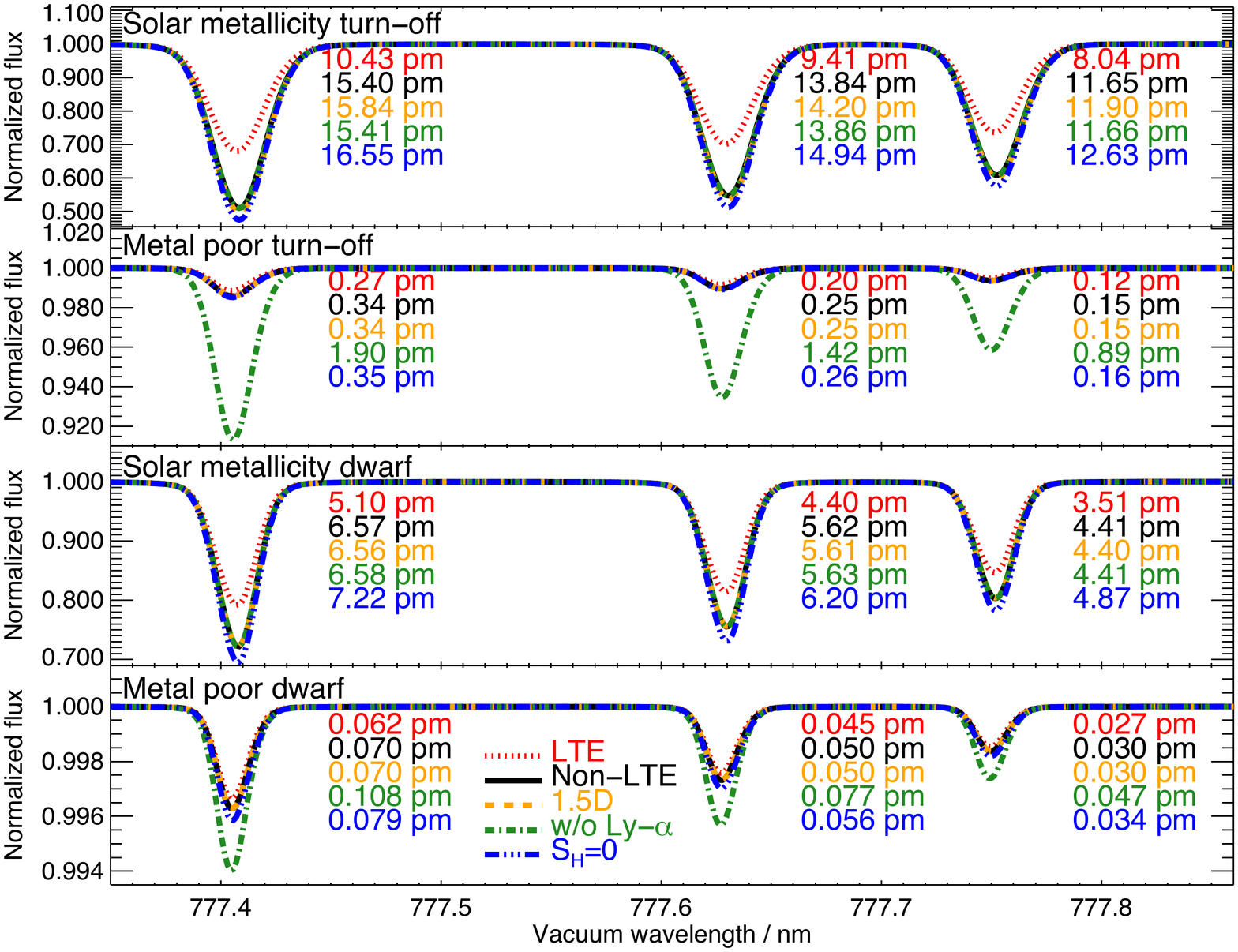}
\caption{Normalized theoretical flux spectra 
(from \mtd~and \scate, as described in \sect{methodgrids})
in the vicinity of the \lna{777}~lines
in a turn-off star ($\teff\approx6500\,\mathrm{K}$, $\lggu=4.0$) 
and in a dwarf star ($\teff\approx5500\,\mathrm{K}$, $\lggu=4.0$) 
with $\feh=0.0$, $\lgeps=8.7$ (solar metallicity)
and $\feh=-3.0$, $\lgeps=6.2$ (metal poor).
Spectra are shown for the 3D LTE, 3D non-LTE,
1.5D non-LTE (\sect{results15d}),
3D non-LTE without Lyman-$\alpha$ (\sect{resultslyman}),
and 3D non-LTE with $\sh=0.0$ (\sect{resultshydrogen});
their equivalent widths in $\mathrm{pm}=10^{-12} \mathrm{m}$,
obtained by direct integration, are indicated in that order
to the right of each line.
1.5D refers to the radiative
transfer scheme used to compute the 
statistical equilibrium; the final
emergent intensities are (in all cases shown)
computed using full 3D radiative transfer.}
\label{spectra}
\end{center}
\end{figure*}

Non-local radiation along oblique rays
from hotter or cooler environments 
can affect the statistical equilibrium
in the line forming regions
\citep[e.g.][]{1977A&amp;A....58..273S,2013A&amp;A...558A..20H}.
The influence of this is gauged by 
repeating the non-LTE calculations 
on the 3D model atmospheres, but
treating the model as an ensemble of 1D columns,
each column effectively 
extended in the horizontal directions
to infinity
\citep[so called 1.5D non-LTE; e.g.][]{1995A&amp;A...302..578K}.

In \fig{15dratio} we show the 
spatially-resolved 
relative differences in the departure coefficients of the
\tripletup~level and of the \tripletlo~level
(i.e.~the upper and lower levels of the \lna{777}~lines) 
between the 1.5D non-LTE case and
the 3D non-LTE case, 
in the same snapshot of the solar metallicity
turn-off star shown in \fig{map1}.
Note that the relative difference in the departure coefficients
are the same as that in the populations,
because the LTE populations (the divisors in \eqn{departurecoefficients})
are a property of the background atmosphere,
which is left unchanged. 

At large depths the populations are close to 
their LTE values in 1.5D and in 3D.
\markcorr{Higher up at $\lgt=-1.5$
(within the peak line forming region),
the differences in the populations of the
\tripletlo~level are of the order $10\%$,
with the populations being larger in the 3D case.
The differences are associated with the reverse granulation.
In 1.5D, the hot reverse granulation features
(thin, bright filaments in the plots)
are infinitely extended in the horizontal directions,
whereas in 3D these features are surrounded by cooler gas.
More photons can be created when the temperature is larger 
(cf. the Planck function).
Consequently, there are fewer non-local photons propagating into 
these hot features in 3D than in 1.5D.
With fewer photoexcitations, the statistical equilibrium in 
these regions is
such that there is a larger overpopulation of \tripletlo~level.
In contrast, the mean radiation field in the surrounding regions 
(dark areas around the thin filaments) is larger in 3D than in 1.5D.
This is because extra non-local photons from the hot regions
propagate into the cooler regions.
With more photoexcitations in these cool regions, 
the \tripletlo~level has a smaller population in 3D than in 1.5D.}
The behaviour of the \tripletup~level is more complicated, 
since there is a flow from the highly excited
levels into it via the \lna{927}~lines, that competes
with the flow out of the level 
via the \lna{777}~lines. 

\markcorr{The statistical equilibrium 
populations are slightly increased or decreased in 3D compared to 1.5D
at different locations in the simulation box.
This means that net effect of non-vertical radiative transfer
on the averaged emergent fluxes is small.
Effects are most severe in solar metallicity 
turn-off stars, in which the \lna{777}~lines form high up 
in the temperature inhomogeneities 
associated with reverse granulation
play a significant role;
\fig{15dratio} is indicative of the worst case.}

We compare in \fig{spectra} theoretical 
disk-integrated flux spectra
in the vicinity of the \lna{777}~lines
for a variety of test cases, including 1.5D non-LTE.
(The final emergent intensities are
computed using full 3D radiative transfer.)
The 1.5D non-LTE line strength is of the order 0.01 dex
larger than the standard 3D non-LTE case
in the most extreme case (solar metallicity turn-off stars).
This suggests the \lna{777}~lines
could be modelled accurately
and at considerably smaller computational cost
by using efficient 1.5D non-LTE radiative transfer codes 
\citep[e.g.][]{2015A&amp;A...574A...3P} 
to compute the statistical equilibrium.

\subsection{Background UV opacity at low \feh}
\label{resultslyman}

\begin{figure}
\begin{center}
\includegraphics[scale=0.31]{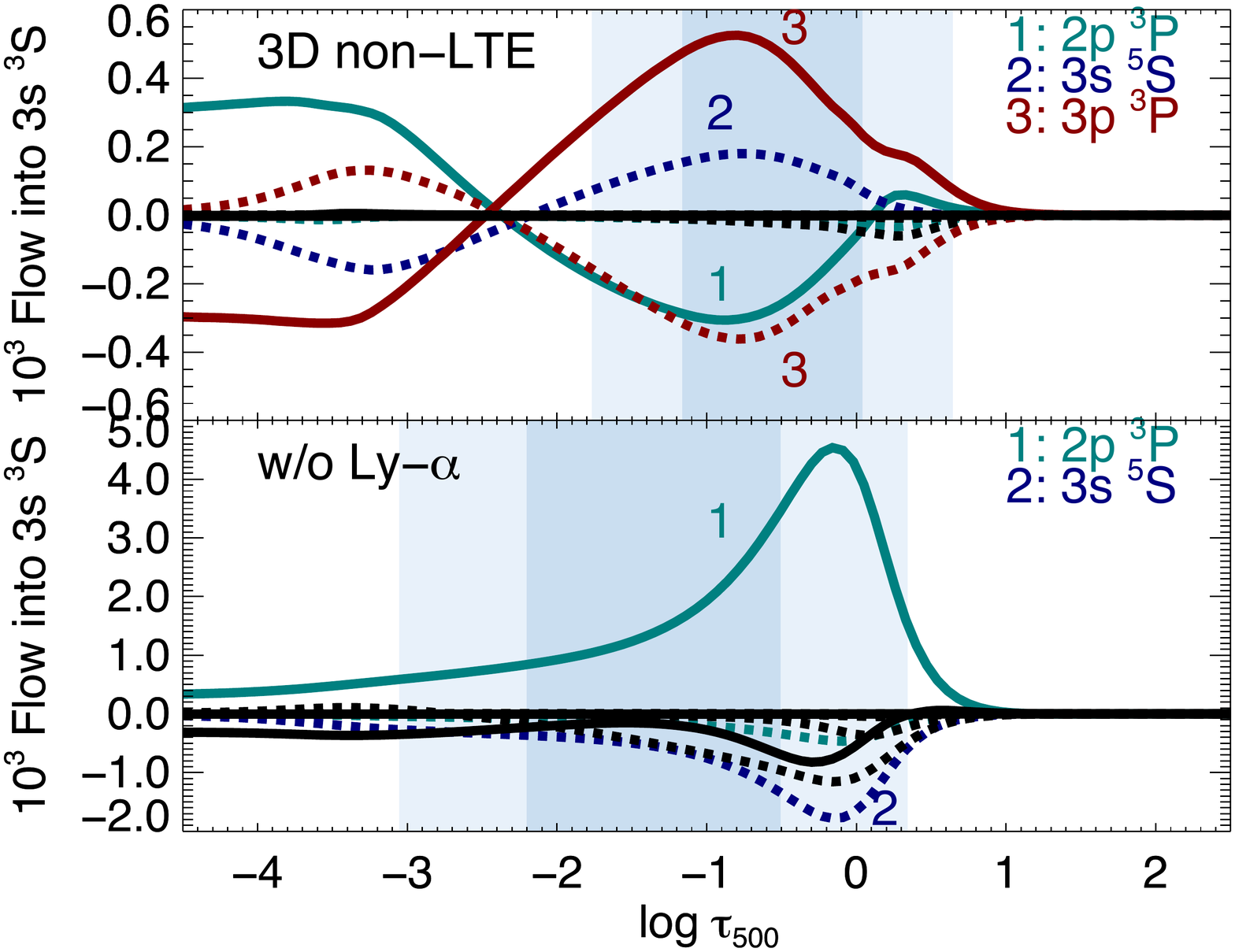}
\caption{Normalized net flow
into the \term{3}{s}{3}{S} level
(the upper level of the \lna{130}~lines)
at different depths
in a metal poor turn-off star 
($\teff\approx5500\,\mathrm{K}$, $\lggu=4.0$,
$\feh=-3.0$, $\lgeps=6.2$)
with and without 
Lyman-$\alpha$ background line opacity
in the non-LTE analysis. 
Collisional transitions are indicated as dashed lines
while radiative transitions are indicated as solid lines.
Plotted here are median values
on surfaces of equal $\lgt$.
The shaded areas correspond to the
mean formation depth $\pm$ one standard deviation,
and $\pm$ two standard deviations
\markcorr{(with the mean formation depth
and its standard deviation defined in \fig{bdd})}.}
\label{bdd2}
\end{center}
\end{figure}

\citet{2009A&amp;A...500.1221F} reported
large departures from LTE at low metallicities,
associated with photon pumping in the 
UV resonance \lna{130}~lines.
However, their analysis did not include 
the $121.5\,\mathrm{nm}$ Lyman-$\alpha$ transition
as a background opacity,
which is significant because the line is very strong and
the \lna{130}~lines sit on its red wing.
Neglecting Lyman-$\alpha$, we were able
to reproduce the large non-LTE effects
in the \lna{777}~lines; when this transition
is included in the analysis however,
the statistical equilibrium at low metallicities
is much closer to LTE.

We compare in \fig{bdd2} the normalized net flow
into the \term{3}{s}{3}{S}~level
(i.e.~the upper level
of the UV resonance \lna{130}~lines)
when Lyman-$\alpha$ is included and 
when it is not.
When Lyman-$\alpha$ is neglected,
the background opacity in the UV 
is significantly lowered.
This allows for an abundance of UV photons
being absorbed via the 
UV resonance \lna{130}~lines,
driving a large overpopulation with respect to LTE
in the \term{3}{s}{3}{S}~level.
(The lower levels of these lines
is the ground state of oxygen,
which retain their LTE populations.) 
As can be seen in \fig{bdd2},
and as discussed by \citet{2009A&amp;A...500.1221F},
collisional coupling between
the \term{3}{s}{3}{S}~level and the \tripletlo~level 
(i.e.~the lower level of the \lna{777}~lines)
propagates a flow from the former level
over to the latter level.
The overpopulation is then 
propagated to the \tripletup~level
(i.e.~the upper level of the \lna{777}~lines)
by collisional coupling.
This increases the \lna{777}~line opacity,
leaving the source function only slightly perturbed.

The Lyman-$\alpha$ line
provides an efficient alternative destruction route
for UV photons that largely stifles the photon pumping effect
in the \lna{130}~lines.
Without the photon pumping in the \lna{130}~lines,
the \lna{777}~line opacity is closer to its LTE value;
in other words we find the statistical equilibrium 
to be much closer to LTE in the metal poor regime.

We compare in \fig{spectra} theoretical flux spectra
in the vicinity of the \lna{777}~lines
when Lyman-$\alpha$ is included and 
when it is not. 
The importance of this effect 
is greatest in metal poor turn-off stars
(second panel of \fig{spectra}),
in which neglecting Lyman-$\alpha$
strengthens the \lna{777}~lines 
by of the order 0.75 dex 
(with the exact value depending on the choice of 
oxygen abundance).
At lower effective temperatures
and higher surface gravities,
and particularly at higher metallicities,
neglecting Lyman-$\alpha$ incurs
a smaller error.

\markcorr{Other lines and continua in the UV providing background opacity 
were included in the analysis, since they can
also play a role in determining the statistical equilibrium
of oxygen in metal poor stars.
For example, the Lyman-$\beta$~line at 102.6nm
was included in the analysis.
It overlaps with a number of other UV oxygen lines;
the mechanism by which the statistical equilibrium is affected 
is similar to that described above for
Lyman-$\alpha$~and the \lna{130}~lines.
We emphasise however that Lyman-$\alpha$~appears to be
the most important UV background opacity to include,
at least for the metal poor stars studied in this work. 
For more discussion on the effects of Lyman-$\beta$~and of 
other background UV opacities on the 
statistical equilibrium we refer the reader to Sect.~3.1.2~of 
\citet[][]{2009A&amp;A...500.1221F}.}

\subsection{Neutral hydrogen collisions}
\label{resultshydrogen}

It is important to consider how
the neutral hydrogen
collisions affect the results,
because the adopted recipes are very uncertain
\citep[e.g.][]{2011A&amp;A...530A..94B}.

The collisional transition with the largest
effect on the \lna{777}~lines is that
between the \tripletup~level and the \tripletlo~level
(i.e.~the upper and lower levels of the lines).
Collisional coupling is responsible for an upwards flow 
between these two levels
that compensates the downwards flow 
driven by the corresponding radiative transition (\fig{bdd}).
Obtaining accurate neutral hydrogen 
collisional rate coefficients 
for this transition are of vital importance.
Of secondary importance are the rate coefficients
for the collisional transitions 
between the highly excited upper levels (\term{4}{s}{5}{S}~and above).
The close coupling between these levels is maintained 
by neutral hydrogen collisions,
rather than electron collisions; 
errors in these rates will influence the downwards
cascade into the \tripletup~level
via the \lna{927}~line
(see \sect{resultsnonlte}).

Fortunately, the LTE abundance errors for the \lna{777}~flux 
do not seem to be too sensitive 
to the hydrogen collisions. 
We show in \fig{spectra} theoretical flux spectra
in the vicinity of the \lna{777}~lines
when $\sh$ is set to zero,
so that the collisions with neutral hydrogen 
are neglected entirely.
In the cases shown, and more generally across the grid,
neglecting hydrogen collisions strengthens
the \lna{777}~lines by of the order 0.05 dex.
\markcorr{It bears repeating that the main results of this work 
are calculated with $\sh$ set to unity
(i.e.~no scaling of the rate coefficients), 
this choice being based on a solar analysis 
of the \lna{777}~lines (\sect{methodatom}).}


\subsection{Abundance errors}
\label{resultsabundanceerrors}

\begin{figure*}
\begin{center}
\includegraphics[scale=0.66]{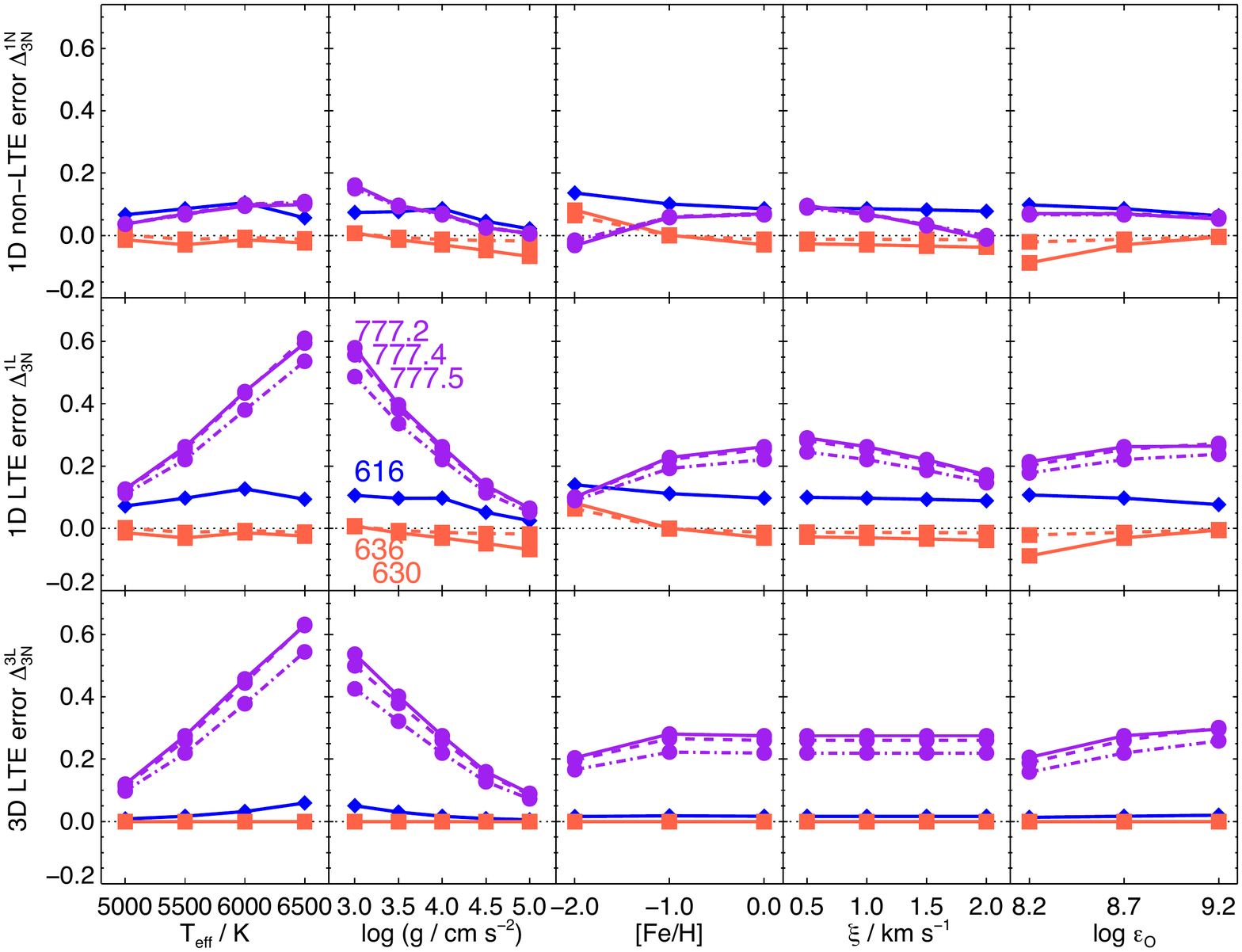}
\caption{1D non-LTE abundance errors ($\corr{1N}{3N}$),
1D LTE abundance errors ($\corr{1L}{3N}$), 
and 3D LTE abundance errors ($\corr{3L}{3N}$)
for various atomic oxygen lines.
The columns show how the abundance errors
depend on a single parameter,
with the other four parameters fixed.
When the parameters are fixed,
they have values 
$\teff=5500\,\mathrm{K}$, 
$\lggu=4.0$,
$\feh=0.0$,
1D microturbulence $\xi=1.0\,\kms$,
and 3D non-LTE oxygen abundance $\lgeps=8.7$.
Note that $\xi$ is only relevant in the first and second rows. 
\markcorr{The \lna{616}~line is denoted by
diamonds connected with solid lines.
The \lnf{630}~and \lnf{636}~lines are denoted by
squares connected with solid and dashed lines, respectively.}
The \lnf{630}~line
includes the NiI blend;
the adopted nickel abundance is
solar scaled: 
$\log\epsilon_{\text{Ni}}=6.20+\feh$
\citep{2015A&amp;A...573A..26S}.
Without the blend, the abundance errors
for this line do not depart
significantly from those for the 
\lnf{636}~line.
\markcorr{The three components of the
\lna{777}~lines are labelled with their
wavelengths in air and are denoted by circles connected with
solid, dashed, and dot-dashed lines,
from blue to red respectively.}}
\label{abdiff}
\end{center}
\end{figure*}

Abundance errors (\sect{methodgrids}) are good indicators 
of the importance (or the lack of importance)
of modelling the system using 3D stellar models
and/or multi-level non-LTE radiative transfer. 
\markcorr{We show the abundance errors
as functions of the different parameters 
($\teff$, $\lgg$, $\feh$, $\xi$ and $\lgeps$) in \fig{abdiff}.
The plots in each column illustrate how 
the abundance errors depend on a single parameter,
while the remaining four are fixed.}
The rows show 1D non-LTE abundance errors ($\corr{1N}{3N}$),
1D LTE abundance errors ($\corr{1L}{3N}$), 
and 3D LTE abundance errors ($\corr{3L}{3N}$).

The abundance errors were 
obtained by linear interpolation in the 
1D non-LTE, 1D LTE, and 3D LTE abundances (respectively). 
The abundance errors also had to be
interpolated linearly in the effective temperature, 
because the effective temperature grid is irregular
(see \sect{methodgrids}).

Importantly, the abundance errors plotted
in \fig{abdiff} are all relative to,
and functions of,
the 3D non-LTE oxygen abundance,
enabling a direct comparison between them.
Furthermore, the 
1D plane-parallel hydrostatic 
\atmo~\citep[][Appendix A]{2013A&amp;A...557A..26M}
model atmospheres
have as far as possible identical input physics 
to the \stagger~models
(\sect{methodmodelatmospheres}).
This enables a fair comparison of the non-LTE effects
in 1D and 3D.

We show the abundance errors for the \lnf{777}~lines,
the \lnf{636}~and \lnf{630}~lines, 
and the \lna{616}~line;
these are discussed in turn.

\subsubsection{\lna{777}~lines}

The \lna{777}~lines are very prone
to non-LTE effects. 
As discussed in \sect{resultsnonlte},
\markcorr{photon losses in}
the high excitation 
\lna{777}~lines drive an overpopulation of the
\tripletlo~level (i.e.~the lower level of these lines)
with respect to LTE.
The cascade effect, 
in which there is a flow 
from the highly excited upper oxygen levels 
into the the \tripletup~level
(i.e.~the upper level of the \lna{777}~lines)
via the \lna{927}~line, can also contribute
to the overpopulation of the \tripletlo~level.
Larger abundance errors
(or more negative abundance corrections)
can therefore be expected where 
the high excitation lines are stronger: 
i.e.~where the effective temperature is larger,
the surface gravity is smaller,
and the oxygen abundance is larger
(\fig{abdiff}: third row; first,
second and fifth columns).
The blue component of these three lines
has the largest intrinsic strength,
\markcorr{so it typically has the largest non-LTE abundance differences.
The red component is the weakest,
so it typically has the smallest non-LTE abundance differences.}

The abundance errors have a mild dependence on
stellar metallicity
(\fig{abdiff}: third column).
The trend reflects mild changes in the
mean temperature stratification of the 3D models
at fixed effective temperature:
as \feh~is decreased, the gas temperature
becomes cooler in the line forming regions,
which results in smaller non-LTE effects. 

The 1D non-LTE and 1D LTE abundance errors 
vary with 1D microturbulence parameter $\xi$
(\fig{abdiff}: first and second rows; fourth column).
When the \lna{777}~lines are strong,
increasing this parameter
increases the equivalent widths
of the lines modelled in 1D, while 
the equivalent widths of the lines modelled in 3D 
are fixed
(the 3D grids being computed without
any analogous microturbulence parameter).
This reduces the 1D non-LTE and 1D LTE abundance errors.
When the \lna{777}~lines are weak, 
they lose their sensitivity to
the microturbulence parameter.

\markcorr{Although the 1D non-LTE abundance errors 
are sensitive to the choice of 1D microturbulence parameter,
they are typically of the order 0.05 to 0.1 dex
(\fig{abdiff}: first row).
In other words, the 3D models exacerbate the 
non-LTE effects; part of the reason is 
the efficient \lna{777}~line formation
in the atmospheric inhomogeneities driving
larger departures from LTE,
as discussed in \sect{resultsinhomogeneities}.
This result has practical implications.
Typically, surveys of oxygen abundances in FGK-type stars
that are based on the \lna{777}~lines
adopt the procedure of correcting 1D LTE abundances
using pre-computed 1D non-LTE abundance corrections 
\citep[e.g.][]{2013ApJ...764...78R,2014A&amp;A...568A..25N}.
Our 1D non-LTE abundance errors suggest that 
this procedure systematically overestimates
the oxygen abundance, albeit by a rather small amount.}

\subsubsection{\lna{616}~line}

The high-excitation \lna{616}~line,
originating between the \term{4}{d}{5}{D}~level
and the \tripletup~level,
suffers only mildly from non-LTE effects;
the largest 3D LTE abundance errors
in our grid are of the order 0.05 dex
(\fig{abdiff}: third row).
Like the high-excitation \lna{777}~lines, the departures from LTE 
are larger when the effective temperature is larger,
the surface gravity is smaller,
and when the oxygen abundance is larger.
The line is more susceptible to 3D effects.
1D abundance errors 
are typically of the order 0.1 to 0.2 dex
(\fig{abdiff}: first and second rows).

\subsubsection{\lnf{630}~and \lnf{636}~lines}

The \lnf{630}~and \lnf{636}~lines
do not suffer from non-LTE effects.
(\fig{abdiff}: third row).
The lower levels of these lines 
is the vastly populated ground level of oxygen,
which retains its LTE population. 
The upper levels of these lines 
are in close collisional coupling 
with the lower levels,
ensuring that they too remain in LTE.

Like the \lna{616}~line, these lines
are more susceptible to 3D effects.
1D abundance errors are most
sensitive to the stellar metallicity,
and tend to be larger in metal poor stars
(\fig{abdiff}: first and second rows; third column;
the most severe case, $\feh=-3.0$, is not \markcorr{shown),
in} which the mean temperature stratification 
deviates more from the 1D counterparts
\citep[e.g.][]{2002A&amp;A...390..235N}.
In such stars, we find 1D abundance errors up to 0.15 dex.

For the \lnf{630}~line we included
the Ni blend \citep[e.g.][]{1978MNRAS.182..249L,2001ApJ...556L..63A,
2003ApJ...584L.107J} in LTE and with a 
solar scaled nickel abundance:
$\log\epsilon_{\text{Ni}}=6.20+\feh$
\citep{2015A&amp;A...573A..26S}.
Any differences in the abundance errors
between the \lnf{630}~and \lnf{636}~lines
can be attributed to the different line strengths
of the Ni blend in 1D and in 3D,
since the same nickel abundance was used 
in each case.
Without the blend, the abundance differences for
the \lnf{630} and \lnf{636}~lines
are indistinguishable from each other.
The abundance corrections converge 
onto each other when oxygen,
rather than nickel, dominates the \lnf{630} feature;
for example, when the oxygen abundance is increased
(\fig{abdiff}: first and second rows; fifth column).

\subsection{Grids of equivalent widths and abundance corrections}
\label{resultsgrids}

\begin{table*}
\begin{center}
\caption{\markcorr{Predicted equivalent widths $W$
for different solution schemes
``3N'' (3D non-LTE), ``3L'' (3D LTE), 
``1N'' (1D non-LTE), and ``1L'' (1D LTE),
different oxygen lines
labelled by one of ``616'', ``630'', ``636'', 
``7772'', ``7774'', ``7775'', ``777'',
and specified by the
wavelength windows in air for direct integration
$\lambda_{0}<\lambda<\lambda_{1}$,
different sets of stellar parameters 
($\teff$, $\lgg$, $\feh$, $\xi$),
and different oxygen abundances $\lgeps$.
``7772'', ``7774'' and ``7775''
refer to the three components of the \lna{777}~lines
from blue to red respectively; 
``777'' refers to all three components together.
The \lnf{630}~line contains the Ni blend
in LTE and with a solar scaled nickel abundance:
$\log\epsilon_{\text{Ni}}=6.20+\feh$
\citep{2015A&amp;A...573A..26S}.
Only the first five rows are shown; the full table is available online.}}
\label{eqwtable}
\begin{tabular}{c c c c c c c c c c}
\hline
Scheme 
& Line
& $\lambda_{0} / \mathrm{nm}$
& $\lambda_{1} / \mathrm{nm}$
& $\teff / \mathrm{K}$
& $\lggu$ 
& $\feh$ 
& $\xi / \kms$ 
& $\lgeps$ 
& $W / \mathrm{pm}$ \\
\hline
\hline
3N & 616 & 615.73 & 615.97 & 4998 & 3.0 & -0.0 & 0.0 & 7.7 & 0.0330844\\ 
\hline
3N & 616 & 615.73 & 615.97 & 4998 & 3.0 & -0.0 & 0.0 & 8.2 & 0.0986540\\ 
\hline
3N & 616 & 615.73 & 615.97 & 4998 & 3.0 & -0.0 & 0.0 & 8.7 & 0.2713988\\ 
\hline
3N & 616 & 615.73 & 615.97 & 4998 & 3.0 & -0.0 & 0.0 & 9.2 & 0.6703377\\ 
\hline
3N & 616 & 615.73 & 615.97 & 4998 & 3.0 & -0.0 & 0.0 & 9.7 & 1.5235444\\ 
\hline
... &... & ... & ... & ... & ... & ... & ... & ... & ...\\ 
\hline
\end{tabular}
\end{center}
\end{table*}

\begin{table*}
\begin{center}
\caption{\markcorr{3D non-LTE abundance corrections to 
1D LTE based results $\corr{3N}{1L}$
(see \sect{methodgrids})
for different oxygen lines,
labelled by one of ``616'', ``630'', ``636'', 
``7772'', ``7774'', ``7775'', ``777'',
and specified by the
wavelength windows in air for direct integration
$\lambda_{0}<\lambda<\lambda_{1}$,
different sets of stellar parameters 
($\teff$, $\lgg$, $\feh$, $\xi$),
and different oxygen abundances $\lgeps$.
``7772'', ``7774'' and ``7775''
refer to the three components of the \lna{777}~lines
from blue to red respectively; 
``777'' refers to all three components together.
The \lnf{630}~line contains the Ni blend
in LTE and with a solar scaled nickel abundance:
$\log\epsilon_{\text{Ni}}=6.20+\feh$
\citep{2015A&amp;A...573A..26S}.
$\xi$~refers to the microturbulence parameter
used in the 1D calculations; we emphasise that 
no microturbulence was used in any of the 3D calculations.
Only the first five rows are shown; the full table is available online.}}
\label{abdifftable}
\begin{tabular}{c c c c c c c c c}
\hline
Line
& $\lambda_{0} / \mathrm{nm}$
& $\lambda_{1} / \mathrm{nm}$
& $\teff / \mathrm{K}$
& $\lggu$ 
& $\feh$ 
& $\xi / \kms$ 
& $\lgeps^{\mathrm{1L}}$
& $\corr{3N}{1L}$\\
\hline
\hline
616 & 615.73 & 615.97 & 4998 & 3.0 & -0.0 & 0.5 & 8.2 & -0.1216693 \\
\hline
616 & 615.73 & 615.97 & 4998 & 3.0 & -0.0 & 0.5 & 8.7 & -0.1153454 \\
\hline
616 & 615.73 & 615.97 & 4998 & 3.0 & -0.0 & 0.5 & 9.2 & -0.1046884 \\
\hline
616 & 615.73 & 615.97 & 4998 & 3.0 & -0.0 & 0.5 & 9.7 & -0.0869795 \\
\hline
616 & 615.73 & 615.97 & 4996 & 3.0 & -1.0 & 0.5 & 7.2 & -0.1380291 \\
\hline
... & ... & ... & ... & ... & ... & ... & ... & ...\\ 
\hline
\end{tabular}
\end{center}
\end{table*}

\markcorr{We present in \tab{eqwtable} predicted equivalent widths
for the \lna{616}, \lnf{630}, \lnf{636}, and \lna{777}~lines,
calculated on the nodes of the grid.
Equivalent widths are shown for the four different solution schemes:
3D non-LTE, 3D LTE, 1D non-LTE, and 1D LTE. 
Abundance differences between any two schemes can be derived 
using the data in \tab{eqwtable}.
For convenience, we provide in \tab{abdifftable}~the
3D non-LTE abundance corrections to 1D LTE based results ($\corr{3N}{1L}$),
which are likely to be of most interest to 
the stellar spectroscopy community.}

\markcorr{The data in \tab{eqwtable}~and \tab{abdifftable}~were 
interpolated and extrapolated
so as to obtain equivalent widths and abundance corrections
on regular grids of stellar parameters and input oxygen abundances.
The regularly-spaced data are available upon request, and can also
be found online\footnote{
\url{http://www.mso.anu.edu.au/~ama51}}$^{,}$\footnote{
\url{http://inspect-stars.com}}.}

\section{Comparison with previous 1D non-LTE studies}
\label{comparison}

3D non-LTE calculations for oxygen
have previously been only performed
for the Sun \citep{1995A&amp;A...302..578K,
2004A&amp;A...417..751A,
2009A&amp;A...508.1403P,
2013MSAIS..24..111P,
2015arXiv150803487S}.
A detailed comparison with 
these papers plus a discussion of the solar 
oxygen abundance 
will be presented in a forthcoming paper.
Here we restrict ourselves to
comparing our 1D non-LTE results 
with those from several recently published studies:
\citet{2003A&amp;A...402..343T,
2007A&amp;A...465..271R,
2009A&amp;A...500.1221F,
2013AstL...39..126S}.
Since these studies provide
1D non-LTE abundance corrections ($\corr{1N}{1L}$)\footnote{
With the exception of \citet{2003A&amp;A...402..343T},
in which $\corr{1L}{1N}$ is provided. We interpolated their grid to
obtain $\corr{1N}{1L}$.}, we compare
abundance corrections, rather than
abundance errors, in this section.

The ingredients of the analyses that are likely to
have the most impact on the results are
the model atom, in particular
the description of electron collisions and hydrogen collisions
\citep[e.g.][]{2000A&amp;A...359.1085P,2011A&amp;A...530A..94B}.
Model atmospheres have a number of
associated uncertainties,
even in the 1D case
\citep[e.g. discussions in][]{1997A&amp;A...318..841C};
therefore calculations over different
families of model atmospheres
can have systematically different outcomes.
Missing background opacities,
(e.g. Lyman-$\alpha$; see \sect{resultslyman})
can also be a source of error. 

\subsection{\citet{2013AstL...39..126S}}

\begin{table*}
\begin{center}
\caption{1D non-LTE abundance corrections
($\corr{1N}{1L}$) 
for the blue component of the \lna{777}~lines
in a solar metallicity turn-off star 
($\teff=6000\,\mathrm{K}$, $\lggu=4.0$)
and a solar metallicity dwarf star
($\teff=5000\,\mathrm{K}$, $\lggu=4.0$)
\markcorr{with $\xi=2.0\,\kms$}
and 1D LTE oxygen abundance $\lgeps=8.83$.
We show results 
from \markcorr{Table 11 of}~\citet{2013AstL...39..126S}
for which the \citet{2007A&amp;A...462..781B}
collisional rate coefficients are not scaled,
\markcorr{results from the $\sh=1.0$~grid
of \citet{2009A&amp;A...500.1221F},
results from the grid described in \citet{2007A&amp;A...465..271R},
and results interpolated from Table 2 of \citet{2003A&amp;A...402..343T}.}
Note that the abundance corrections of
\citet{2009A&amp;A...500.1221F}
and \citet{2007A&amp;A...465..271R}
have no nominal dependence on $\xi$.}
\label{abcortable}
\begin{tabular}{c c c c c c}
\hline
Star & This work & {\citet{2013AstL...39..126S}} & {\citet{2009A&amp;A...500.1221F}} & {\citet{2007A&amp;A...465..271R}} & {\citet{2003A&amp;A...402..343T}} \\
\hline
\hline
Turn-off & -0.31 & -0.34 & -0.35 & -0.26 & -0.19 \\
\hline
Dwarf & -0.09 & -0.10 & -0.11 & -0.11 & -0.06 \\
\hline
\end{tabular}
\end{center}
\end{table*}

The authors of this theoretical study performed calculations on
\atlasnine~model atmospheres \citep{1993KurCD..13.....K,1993ASPC...44...87K},
and used \detail~\cite{butler1985newsletter} to solve
the statistical equilibrium. 
The authors updated the
model atom of \cite{2000A&amp;A...359.1085P},
(which includes 51 OI levels plus the OII ground state,
and all optically allowed bound-bound transitions)
to use quantum mechanical 
electron excitation rate coefficients
\citep{2007A&amp;A...462..781B},
and to include neutral hydrogen collisions
following the recipe of
\citet{1968ZPhy..211..404D,1969ZPhy..225..483D} 
as formulated by \citet{1984A&amp;A...130..319S}
(and with $\sh=1.0$);
the treatment of collisional rates
for radiatively-forbidden couplings 
is not discussed.

Included in their study is a small grid of
1D non-LTE abundance corrections
($\corr{1N}{1L}$) 
for the blue component
of the \lna{777}~lines \citep[Table 11 of][]{2013AstL...39..126S}.
Two of their calculations lie within our grid (\tab{abcortable});
for these two cases we find
good agreement with our corresponding abundance corrections.

\subsection{\citet{2009A&amp;A...500.1221F}}

\begin{figure}
\begin{center}
\includegraphics[scale=0.31]{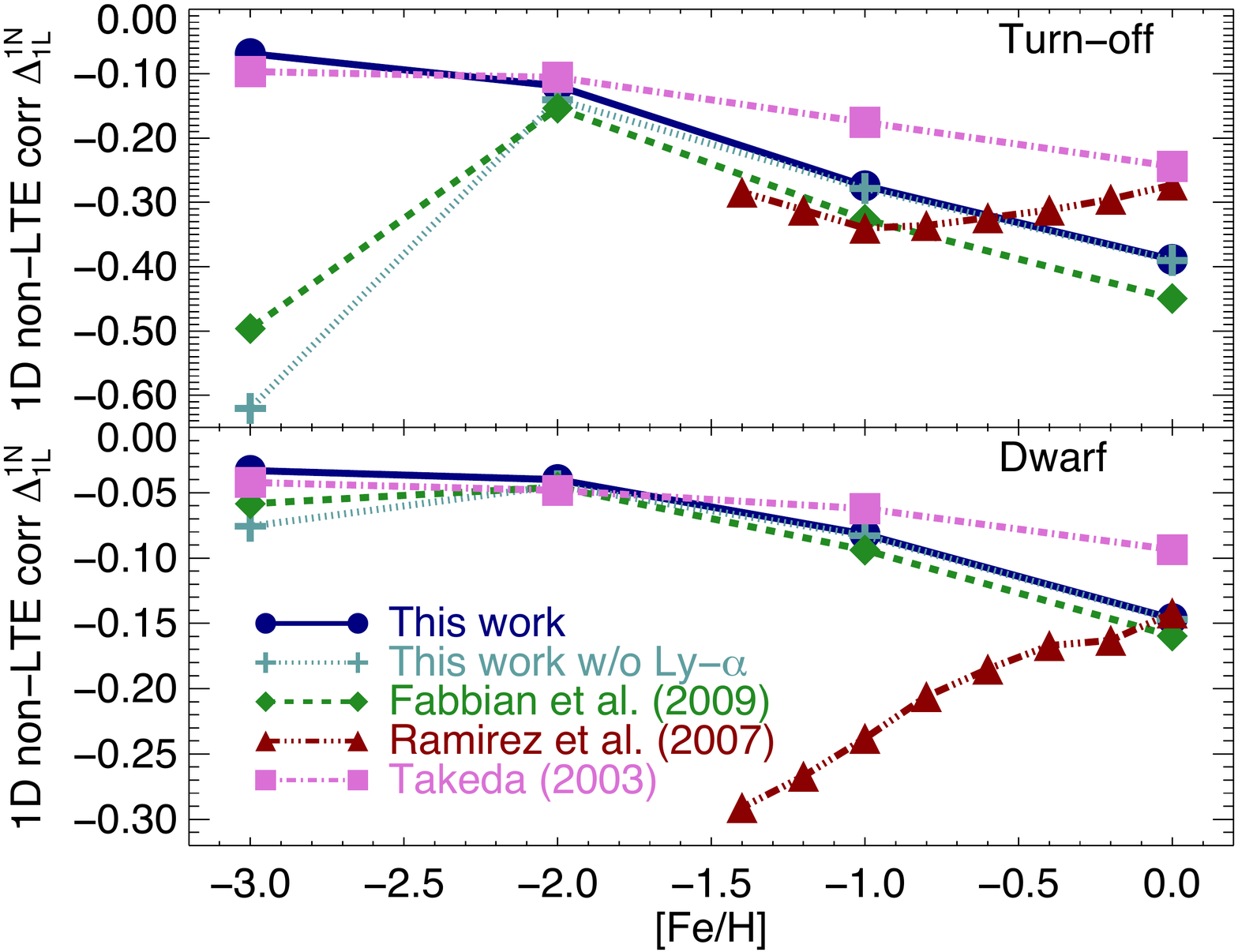}
\caption{1D non-LTE abundance corrections
($\corr{1N}{1L}$) 
for the middle \lna{777}~line,
predicted by this work, 
\citet{2009A&amp;A...500.1221F},
\citet{2003A&amp;A...402..343T},
and \citet{2007A&amp;A...465..271R}.
Also shown are
the abundance corrections predicted by this work 
when background opacity from
Lyman-$\alpha$ is not included.
Top panel is for
a turn-off star ($\teff=6500\,\mathrm{K}$, $\lggu=4.0$),
and the bottom panel is for 
a dwarf star ($\teff=5500\,\mathrm{K}$, $\lggu=4.0$),
with 1D microturbulence
$\xi=1.0\,\kms$ and varying chemical composition.
The abundance corrections are functions of a
measured LTE oxygen abundance
which has a value of $8.7+\feh+[\alpha/\text{Fe}]$, 
where the enhancement
is $[\alpha/\text{Fe}]=0.5$ for $\feh\leq-1.0$,
$[\alpha/\text{Fe}]=0.0$ for $\feh=0.0$,
and a linear function in the region between them
(i.e.~$[\alpha/\text{Fe}]=-0.5\times\feh$ for $-1.0<\feh<0.0$).}
\label{fabbian}
\end{center}
\end{figure}

The authors performed
calculations on \marcs~model atmospheres
\citep[][and subsequent 
updates, prior to the
most recent version 
of \citet{2008A&amp;A...486..951G}]{1975A&amp;A....42..407G},
and \multi~\citep{1986UppOR..33.....C},
version 2.2, to solve
the statistical equilibrium.
This code has a number of features in common with
the code we used (\mtd); in particular
the continuous opacity package is
essentially the same.
Their model atom (53 OI energy levels plus 
the ground state of OII, and 208 bound-bound
transitions)
is larger than that used here,
but uses the same description for electron
collisional rate coefficients 
\citep{2007A&amp;A...462..781B},
and neutral
hydrogen collisional rate coefficients
\cite{1993PhST...47..186L}.
They perform calculations with $\sh=0.0$
and with $\sh=1.0$; 
only the latter case is discussed here.
In particular, in this latter case
the authors use the same value
of $f_{\text{min}}=10^{-3}$.

We show in \fig{fabbian} their 
1D non-LTE abundance corrections
($\corr{1N}{1L}$) 
for the middle $\lna{777}$~line
\markcorr{with $\sh=1.0$},
in turn-off stars and dwarf stars
with varying chemical compositions.
At $\feh\gtrsim-2.0$,
their abundance corrections and our own
are in good agreement.
The turn-off star at $\feh=0.0$ represents
the most discrepant case, where the 1D non-LTE
abundance corrections differ by about $0.06$ dex.
The small discrepancies may be associated with 
differences between the 
\marcs~model atmospheres
and the \atmo~model atmospheres..

At $\feh\lesssim-2.0$, the difference
between our results and theirs is
very large, and is due to the 
background UV opacity effect we discussed in \sect{resultslyman}.
Their analysis did not include the Lyman-$\alpha$ transition
as a background opacity.
We re-ran our 1D non-LTE calculations
with Lyman-$\alpha$ omitted.
As seen in \fig{fabbian}, the agreement
between the two studies
significantly improves,
and the trend with $\feh$ is reproduced.
This strongly suggests that the 
very negative abundance corrections found by 
\citet{2009A&amp;A...500.1221F} at low $\feh$
would be quenched by including background UV 
opacity from the Lyman-$\alpha$ transition.

\subsection{\citet{2007A&amp;A...465..271R}}

The authors performed calculations on 
\atlasnine~model atmospheres \citep{1993KurCD..13.....K,1993ASPC...44...87K},
used \tlusty \citep{1988CoPhC..52..103H}
to solve the statistical equilibrium
and \synspec \citep{1985BAICz..36..214H}
to generate emergent spectra.
Their model atom was that of \citet{2003ApJS..147..363A}
(54 OI levels plus the ground
state of OII, and 242 bound-bound transitions).
This study being antecedent to
the quantum mechanical 
calculations of \citet{2007A&amp;A...462..781B},
the authors used
the classical prescriptions of \citet{1962ApJ...136..906V}
and \citet{1962amp..conf..375S},
for radiatively allowed and forbidden
electron excitation rate coefficients, respectively.
The authors neglected neutral hydrogen collisions.
Instead they apply a small empirical
correction to their results,
independent of stellar parameters,
making their 
1D non-LTE abundance corrections
($\corr{1N}{1L}$) 
less negative by 0.0355 dex and 0.018 dex
for the blue component and the middle
component of the \lna{777}~triplet lines.

Their grid of abundance corrections is
finely spaced in metallicity but only
extends down to $\feh=-1.4$.
We compare their abundance corrections
with our own in \tab{abcortable} and \fig{fabbian}.
The agreement is generally poor.
The trend of their abundance corrections
with $\feh$ is not qualitatively consistent with the 
trends from other studies shown on that plot.
There is a strong sensitivity 
to the stellar metallicity
that we do not find in our non-LTE results.

This discrepancy likely arises from
more than one place.
We note that missing background continuum
opacities and background line opacities in the UV
could drive photon pumping in the
\lna{130}~lines and lead to large 
non-LTE effects in the system
(cf.~the mechanism described in \sect{resultslyman});
this would also explain the
strong dependence on 
the stellar metallicity.

\subsection{\citet{2003A&amp;A...402..343T}}

A detailed description
of their simulation setup can be found in
\citet{1991A&amp;A...242..455T,1992PASJ...44..309T,
2003A&amp;A...402..343T}. 
The author performed calculations on \atlasnine~models 
\citep{1993KurCD..13.....K,1993ASPC...44...87K},
using a large model atom (87 OI levels
and 277 bound-bound transitions).
The author used
the classical prescriptions of \citet{1962ApJ...136..906V}
and \citet{1973ApJ...184..151A}
for the radiatively-allowed
and radiatively-forbidden
electron excitation rate coefficients.
For the analogous neutral hydrogen 
rate coefficients they 
used the recipe of
\citet{1968ZPhy..211..404D,1969ZPhy..225..483D} 
as formulated by \citet{1984A&amp;A...130..319S}
(and with $\sh=1.0$),
and modified the formula in \citet{1973ApJ...184..151A}
for the radiatively-forbidden couplings.

Using their Table 2, we compare \citet{2003A&amp;A...402..343T}
1D non-LTE abundance corrections
($\corr{1N}{1L}$) 
with our own in \tab{abcortable} and \fig{fabbian}.
In contrast to what was found 
with \citet{2009A&amp;A...500.1221F},
we find very good agreement towards lower metallicities,
and less good agreement at solar metallicity,
with a discrepancy in the latter case that can be upwards of 0.15 dex.
The good agreement at low metallicity 
reflects the fact that the non-LTE effects 
in the $\lna{777}$~lines
are driven by the strength of the lines themselves,
so as they get weaker the departures
from LTE must get less severe.
We note that \citet{2003A&amp;A...402..343T}
have included the Lyman-$\alpha$ line in their analysis
\citep[see the discussion in \S4.3 of][]{2009A&amp;A...500.1221F}.

The discrepancies at higher metallicities
are likely because of the different
electron collisional rate coefficients used.
As demonstrated by
\citet{2009A&amp;A...500.1221F} and
\citet{2013AstL...39..126S},
the new rate coefficients by \citet{2007A&amp;A...462..781B}
drive larger departures from LTE.

\section{Conclusions}
\label{conclusion}

We have studied non-LTE atomic oxygen line formation
across a grid of theoretical 3D hydrodynamic \stagger~model
atmospheres of turn-off stars, dwarfs and subgiants. 
The main findings are as follows.
\begin{itemize}

\item{Non-LTE effects on the \lna{777}~lines are mostly driven by 
\markcorr{photon losses in} the lines
themselves, which means their impact are largest when the lines themselves 
are stronger (i.e.~at larger effective temperatures,
smaller surface gravities, and larger oxygen abundances).
1D LTE abundances and 3D LTE abundances 
are over 0.2 dex too large, reaching 0.6 dex or
more for solar metallicity turn-off stars.}

\item{3D effects (mean temperature stratification,
atmospheric inhomogeneities) exacerbate the non-LTE effects.
Consequently, 1D non-LTE abundances
can be of the order \markcorr{0.05 to 0.1 dex} too large, with the
exact error depending on the adopted 1D microturbulence.}

\item{Non-vertical radiative transfer (another 3D effect)
changes the \lna{777}~disk-integrated line strengths
by only 0.01 dex. This suggests that these lines can be modelled
accurately in 1.5D non-LTE.}

\item{It is crucial to include all UV background opacities.
Neglecting the Lyman-$\alpha$ transition introduces a spurious oxygen
non-LTE mechanism (photon pumping in the \lna{130}~UV resonance lines) 
which increases the \lna{777}~line strengths by up to 0.75 dex in 
metal poor turn-off stars.}

\item{The treatment of neutral hydrogen collisions remains a deep 
uncertainty in the study of oxygen line formation. 
Neglecting hydrogen collisions increases the non-LTE \lna{777}~line strengths
by of the order 0.05 dex.}

\item{The \lna{616}~line suffers mildly from non-LTE effects,
and the \lnf{630}~and \lnf{636}~lines do not 
show any significant departures from LTE.
These lines however should be modelled in 3D,
with 1D abundance errors of the order 0.2 dex
in the worst cases.}

\end{itemize}

We have provided predicted
3D non-LTE based equivalent widths
and abundance corrections to 1D LTE based results  
for the \lna{616}, \lnf{630}, \lnf{636}~lines, and \lna{777}~lines
\markcorr{(\sect{resultsgrids}: \tab{eqwtable} and \tab{abdifftable}).
These results should be useful to spectroscopic surveys of FGK-type stars.}

\section*{Acknowledgements}
\label{acknowledgements}
We thank Ivan Ramírez for
providing his abundance differences
and interpolation routines,
which were used in \tab{abcortable} and \fig{fabbian},
\markcorr{and Karin Lind for helping with uploading
the results onto the online non-LTE database INSPECT.}
AMA and MA are supported by the Australian
Research Council (ARC) grant FL110100012.
RC acknowledges support from the ARC through DECRA grant DE120102940.
This research was undertaken with the 
assistance of resources from the 
National Computational Infrastructure (NCI),
which is supported by the Australian Government.


\bibliographystyle{mn2e}
\bibliography{/Users/ama51/Documents/work/papers/allpapers/bibl.bib}

\label{lastpage}
\end{document}